\documentclass[12pt,preprint]{aastex}

\usepackage{natbib}
\usepackage{graphics}
\usepackage{epsfig}
\usepackage{amssymb}

\shorttitle{OBSERVING THE RATE OF TIDAL SYNCHRONIZATION}
\shortauthors{Meibom et al.}

\begin{document}

\title{AN OBSERVATIONAL STUDY OF TIDAL SYNCHRONIZATION IN SOLAR-TYPE BINARY
STARS IN THE OPEN CLUSTERS M35 AND M34
\altaffilmark{1}}

\author{S{\o}ren Meibom\altaffilmark{2,3,4} and
Robert D. Mathieu\altaffilmark{4}}
\affil{Astronomy Department, University of Wisconsin - Madison,
    Madison, WI - 53706, USA}

\author{Keivan G. Stassun\altaffilmark{4}}
\affil{Physics and Astronomy Department, Vanderbilt University,
    Nashville, TN - 32735, USA}

\altaffiltext{1}{WIYN Open Cluster Study. XXIV.}
\altaffiltext{2}{{\it smeibom@cfa.harvard.edu}}
\altaffiltext{3}{Current affiliation: Harvard-Smithsonian Center
for Astrophysics, 60 Garden Street, Cambridge, MA - 02138, USA}
\altaffiltext{4}{Visiting Astronomer, Kitt Peak National Observatory,
National Optical Astronomy Observatory, which is operated by the
Association of Universities for Research in Astronomy, Inc. (AURA)
under cooperative agreement with the National Science Foundation.}


\begin{abstract} \label{abs}

We present rotation periods for the solar-type primary stars
in 13 close ($a \la 5$ AU) single-lined spectroscopic binaries
with known orbital periods ($P$) and eccentricities ($e$).
All binaries are members of the open clusters M35 (NGC2168;
$\sim$ 150 Myr) and M34 (NGC1039; $\sim$ 250 Myr). The binary
orbital parameters and the rotation periods of the primary stars
were determined from time-series spectroscopy and time-series
photometry, respectively. Knowledge of the ages, orbital periods,
and eccentricities of these binaries combined with the rotation
periods and masses of their primary stars makes them particularly
interesting systems for studying the rates of tidal circularization
and synchronization. Our sample of 13 binaries includes six with
orbital periods shortward of 13 days ($a \la 0.12$ AU). The stars
in these binaries orbit sufficiently close that their spins and
orbits have evolved toward synchronization and circularization
due to tidal interactions. We investigate the degree of tidal
synchronization in each binary by comparing the angular rotation
velocity of the primary stars ($\Omega_{\star}$) to the angular
velocity expected if the primary star was synchronized ($e = 0$)
or pseudo­synchronized ($e > 0$) with the orbital motion ($\Omega_{ps}$).
Of the six closest binaries two with circular orbits are not
synchronized, one being subsynchronous and one being supersynchronous,
and the primary stars in two binaries with eccentric orbits
are rotating more slowly than pseudosynchronism. The remaining
two binaries have reached the equilibrium state of both a circularized
orbit and synchronized rotation. As a set, the six binaries present
a challenging case study for tidal evolution theory, which in
particular does not predict subsynchronous rotation in such close
systems.

\end{abstract}



\section{INTRODUCTION}	\label{intro}

Tidal and dissipative forces in close detached binary stars
drive an exchange of angular momentum between the rotation
of the stars and their orbital motion. The cumulative effects
of such tidal interactions with time is referred to as tidal
evolution. The characteristic signs of tidal evolution are: 1)
Alignment of the stellar spin axes perpendicular to the orbital
plane; 2) Synchronization of the rotation of the stars to the
orbital motion; 3) Circularization of the orbits. In a population
of coeval detached late-type binary stars tidal evolution can
be observed among the closest binaries ($a \la 0.2$ AU).

Observations of {\it tidal circularization} have been the
primary constraint on tidal theory over the past two decades
\citep[e.g.][]{mm05,mmd04,lst+02,mca+01,dmm92,mdl+92}.
In particular, the distribution of orbital eccentricities
with orbital periods (the $e-\log(P)$ diagram) has provided
clear evidence for tidal circularization in homogeneous
and coeval populations of late-type binaries, and enabled a robust
measure of the degree of circularization integrated over
the lifetime of the binary population as a function of
orbital period. \citet{mm05} define the {\it tidal circularization 
period} of a binary population as the longest orbital period
to which binaries with initial eccentricities of e=0.35
circularize at the age of the population. Importantly, current
theories of tidal circularization cannot account for the
distribution of tidal circularization periods with population
age.

In comparison, the amount of observational data suitable for
measuring the rate of {\it tidal synchronization} in late-type
binaries is sparse. This is in part because binary orbital elements
are simpler to obtain than rotation periods for stars in binaries.
Even so, observations of the synchronization of the stars in a
binary system provide a second window on the tidal effects on
that system, and one of major importance for several reasons.
First, tidal theory makes explicit predictions for the relative
rates of tidal circularization and synchronization that can be
straightforwardly tested \citep{ws02,zb89}. For a typical binary
the two rates differ and the predicted evolutionary paths from
an asynchronous, eccentric binary to a synchronized and circular
binary take the stars in and out of synchronism during the
evolutions of the stars, their orbital separation, and their
orbital eccentricity (see Figure 1 in \citet{zb89} and Figures
1 and 3 in \citet{ws02}).
Second, the range in binary separations over which tidal
synchronization can significantly affect the stellar angular
momentum evolution provides an important constraint on the
impact of binarity on stellar angular momentum evolution
in late-type stars.
Third, synchronization of the observable surface layers is
closely linked to the internal angular momentum transport
in the star. Thus, the rate of synchronization of the surface
layers can shed light on the coupling between those layers
and the stellar interior.
Fourth, observations of magnetic field tracers (X-rays, chromospheric
emission, etc.) during rapid stellar evolution in tidally synchronized
binaries may enable determination of the evolution rates of stellar
dynamos.

Finally, significant tidal synchronization and circularization
is expected in the many star-planet systems where the planets
orbits very close to their late-type host stars (i.e., ``hot Jupiter''
systems). Tidal interactions in star-planet systems may therefore
play an important role in determining the observed distributions
of mass, orbital period, and eccentricity of extra-solar planets
\citep[e.g.][]{ol04}.

To date, most published studies of synchronization have focused
on early-type binaries \citep[e.g.][]{ab04,alg02,gmm84a,gmm84c,levato74}.
This emphasis is due, in part, to telescope and instrument capabilities
which in the past have favored bright and rapidly rotating stars.
Also, the use of archived data on eclipsing binaries has introduced
a bias toward higher mass stars.

However, a few recent studies of tidal synchronization, e.g
\citet{gmm84d}, \citet{cgc95} and \citet{cc97}, include binaries
with late-type main-sequence primary stars. These studies represent
important contributions to the study of tidal synchronization.
\citeauthor{gmm84d} studied 43 detached double-lined eclipsing
and non-eclipsing binaries with at least one late-type stellar
component. They found that the observed degree of synchronism
was in agreement with the theoretical predictions of \citet{zahn77}.
In half of the binaries in their study the primary or secondary star,
or both, have evolved off the main sequence and stellar rotation
was derived from either line-broadening ($v\sin(i)$) or from periodic
brightness variations. For the non-eclipsing binaries in their study,
the stellar rotation velocities measured from line broadening are
less suitable for constraining models of tidal synchronization,
because of the ambiguities introduced by the unknown inclination
of the rotation axis and non-rotational line-broadening due to
the secondary spectrum.

\citeauthor{cgc95} and \citeauthor{cc97} studied tidal
synchronization using eclipsing binary data from \citet{a91},
of which 10 systems have late-type stellar components.
The two studies compared the observed binary parameters
against the model predictions of \citet{tassoul87,tassoul88}
and \citet{zahn89}, respectively, and found general agreement
between the observed level of synchronization and the predicted
timescales for synchronization. Astrophysical parameters
required to determine the theoretical time-scale for which
synchronization is achieved were obtained by comparing theoretical
models \citep{claret95,cg95} directly to each star of the Andersen
sample. Rotation velocities ($v\sin(i)$) for all stars were
derived from line-broadening.

Knowledge of binary ages is critical in order to measure the
rate and evolution of tidal synchronization.  Furthermore, because
of the sensitive dependence of tidal effects to stellar radius,
knowledge of the stellar evolutionary state and history is essential
for testing tidal evolution. For example, tidal theory predicts that
the pre main-sequence (PMS) and the early main-sequence ($t \la 500$
Myr) are the most active phases of tidal evolution. The PMS phase
because of the large radii and deep convective envelopes, and the
early main-sequence phase because the stars are spinning super-synchronously
after contracting onto the zero-age main sequence. Similarly,
post-main-sequence evolution of one or both components of a binary
greatly increases the rate of tidal evolution.

Thus an optimal binary sample for the study of tidal evolution
would comprise a coeval population of late-type binaries with
accurate information about age, evolutionary stage, orbital
parameters, and rotational angular velocities of the stars.
Arguably such binary samples with ages $t \la 500$ Myr are
particularly interesting. Given these needs, young open clusters
are superb laboratories for the study of tidal evolution.

So motivated, we have undertaken parallel spectroscopic and
photometric surveys of the open clusters M35 ($\alpha_{2000}
= 6^{h}~9^{m}$, $\delta_{2000} = 24\degr~20\arcmin$) and M34
($\alpha_{2000} = 2^{h}~42^{m}$, $\delta_{2000} = 42\degr~46\arcmin$)
to derive orbital periods and eccentricities as well as stellar
rotation periods for late-type binary stars. The goal of this
study is to measure the degree of tidal synchronization of the
primary stars in the closest binaries ($a \la 0.2$ AU) and to
compare the results to the predictions from tidal theory.

M35 \citep[$150$ Myr;][]{vss+02,deliyannis06} and M34
\citep[$250$ Myr;][]{steinhauer06} provide populations of close
late-type binaries with ages during the most active phase
of tidal evolution, making them attractive targets for observational
testing of models of tidal synchronization. M35, in particular, provides
a rich population of close binaries which has allowed determination
of a well-defined tidal circularization period at $10.2^{+1.0}_{-1.5}$
days \citep{mm05}. Indeed, eight out of the nine M35 binaries with periods
less than $\sim$ 10 days have been circularized to eccentricities
less than 0.05. In M34, our ongoing spectroscopic survey has
led to the discovery of five circular binaries ($e < 0.1$) with
periods of less than 5.5 days. A tidal circularization period
has not yet been determined for M34.

We begin by briefly introducing the current theories of tidal
evolution in Section~\ref{theory}. Section~\ref{obs}
outlines our observational program and Section~\ref{results}
describe our observational results. In Section~\ref{effects}
we address potential complications of measuring the rotation
periods of stars in close binary systems. In Section~\ref{loop}
we evaluate the degree of tidal synchronization and circularization
for the closest binaries in M35 and M34 and introduce the
$\log(\Omega_{\star}/\Omega_{ps})-\log(P)$ diagram which
presents the dependence on stellar separation. We compare
the observed tidal evolution of individual binaries to the
predictions of current tidal theory in Section~\ref{discussion}.
Section~\ref{conclusions} summarizes and presents our conclusions.


\section{MODEL PREDICTIONS OF TIDAL SYNCHRONIZATION} \label{theory}

We will discuss our observational results in the context of
theoretical models of tidal evolution in solar-type binaries.
The equilibrium tide theory \citep{zahn89,hut81,zahn77} has
been the primary theory used to explain tidal evolution in
main-sequence binaries with late-type components and was
extended by \citet{zb89} to include tidal evolution during
PMS evolution. The physical mechanism responsible for tidal
dissipation is turbulent viscosity in the outer convective
layers of binary component stars. Alternatively, the dynamical
tide theory \citep{zahn77,zahn75} which before 1998 was used
primarily to explain tidal evolution in binaries with early-type
stars, has recently been applied to binaries with solar-type
components \citep{ws02,sw02,tpn+98,gd98}. In the dynamical tide
theory tidally induced internal gravity modes are thermally
damped and dissipated in the convective envelope.

In the equilibrium tide theory the characteristic times for
tidal synchronization and circularization are \citep{zahn89}

\begin{equation}
t_{sync} = \frac{t_{diss}}{6~\lambda_{sync}~q^2}~\frac{I}{M~R^2}~
\Big(\frac{a}{R}\Big)^6
\end{equation}

\begin{equation}
t_{circ} = \frac{t_{diss}}{21~\lambda_{circ}~q(1+q)}~\Big(\frac{a}{R}\Big)^8.
\end{equation}

\noindent Both times are strongly dependent on the ratio of the
stellar separation ($a$) to the stellar radius ($R$) and thus
restrict significant tidal evolution to the closest binaries.
$M$ and $I$ denote the mass and moment of inertia of the primary
star, $q$ the binary mass ratio, $t_{diss}$ the viscous dissipation
time, and $\lambda_{sync/circ}$ a structural constant whose value
depends on the mass concentration within the stars and on where
in the star the tidal torque is applied.

From the ratio of eqs. [1] and [2] it can be estimated that
$t_{circ} \simeq 10^3 \times t_{sync}$ for a binary with solar-type
components. The process of tidal synchronization thus proceeds much
faster than tidal circularization. This difference in timescales
is primarily because the angular momenta of the individual stars
($\sim I \Omega << M R^2 \Omega$) are much smaller than that of the
orbit ($\sim M a^2 \omega$). Here $\omega$ and $\Omega$ are the
orbital angular velocity and the stellar rotation angular velocity,
respectively. A similar difference in the timescales for tidal
circularization and synchronization is found in the dynamical
tide theory (see e.g. eqs. [41] and [42] in \citet{tpn+98}).
Based on the finding from theory of a much shorter timescale for
tidal synchronization than for circularization, it is expected
that, at times when stellar structure is roughly constant over
tidal evolution timescales (such as after the ZAMS), eccentric
binaries will come to synchronization faster than circularization.

During this process, the stars in an eccentric binary will first
synchronize their spin angular velocities to a value $\Omega_{ps}$,
close to (within $\sim$ 20\%) the orbital angular velocity at periastron
passage ($\omega_{p}$) where the stellar separation is minimum.
This is referred to as pseudo-synchronization. An expression for
$\Omega_{ps}$ can be derived by setting the tidal torque on the
stars integrated over the eccentric orbit equal to zero. In the
weak friction (constant time-lag) approximation \citep[see][]{hut81},

\begin{equation}
\Omega_{ps}~~=~~
\frac{1 + \frac{15}{2}e^2 + \frac{45}{8}e^4 + \frac{5}{16}e^6}
{(1 + e)^2~(1 + 3e^2 + \frac{3}{8}e^4)}~~\omega_{p}
\end{equation}

\begin{equation}
\omega_{p}~~=~~\omega~~\frac{(1 + e)^2}{(1 - e^2)^{3/2}}.
\end{equation}

\noindent Thus, again when stellar structure is roughly constant,
it is expected from tidal theory that the rotation of a star in an
eccentric binary should be pseudo-synchronized ($\Omega_{\star} =
\Omega_{ps}$) if the orbital period is similar to or shorter than
the tidal circularization period.

More specific predictions for the evolution of tidal synchronization
and circularization have been made in \citet{zb89} (equilibrium tide
theory) and \citet{ws02} (dynamical tide theory) assuming specified
sets of initial stellar and binary conditions. A detailed graphical
illustration of the tidal evolution of a binary with two $1.0~M_{\odot}$
stars is given by \citet{zb89}. In their model, synchronization and
circularization are achieved in less than 100,000 years due to large
radii and deep convection during the PMS phase. The stars
then spin up due to less efficient tidal braking as the convection
retreats and the stars contract onto the ZAMS. Thus at ages comparable
to M35 and M34, \citeauthor{zb89} predict supersynchronous rotation
in circularized binaries at the ZAMS. Once the stars have settled on
the main sequence, synchronization resumes and is completed by an age
of $\sim$ 1 Gyr. \citeauthor{ws02} present in their Figures 1 and 3
the tidal evolution of a binary with two $1.0~M_{\odot}$ stars in the
framework of the dynamical tide theory. In their model, starting at
the ZAMS, pseudo-synchronization is gradually achieved within $\sim$
500 Myr.

The observational data presented in this paper provide orbital periods
and eccentricities as well as stellar rotation periods for 13 main-sequence
binaries with known ages. We are thus equipped to compare our observations
of tidal evolution to the predictions derived from the findings of
tidal theory.


\section{OBSERVATIONS AND DATA REDUCTION} \label{obs}

We have conducted two parallel observational programs on
the open clusters M35 and M34: 1) High precision radial-velocity
surveys to identify binaries and determine their orbital parameters;
2) Comprehensive photometric time-series surveys to
determine stellar rotation periods from light modulation
by star-spots on the surfaces of the late-type primary stars.

\subsection{Time-Series Spectroscopy}

M35 and M34 have been included in the WIYN Open Cluster Study
\citep[WOCS;][]{mathieu00} since 1997 and 2001, respectively.
As part of WOCS the solar-type stars in both clusters have been
targets in extensive radial-velocity surveys to determine cluster
membership and to detect binary stars. A detailed description
of the radial-velocity surveys of these two clusters will follow
in later papers; we give here the most relevant information.

All spectroscopic data were obtained using the WIYN \footnote{The
WIYN Observatory is a joint facility of the University of
Wisconsin-Madison, Indiana University, Yale University, and the
National Optical Astronomy Observatories.} 3.5m telescope
at Kitt Peak, Arizona, USA. The telescope is equipped with a
Multi-Object Spectrograph (MOS) consisting of a fiber optic
positioner (Hydra) feeding a bench mounted spectrograph. The
Hydra positioner is capable of placing $\sim$ 95 fibers in a
1-degree diameter field with a precision of $0.2\arcsec$. In
the field of M35 and M34 approximately 82-85 fibers are positioned
on stars while the remaining fibers are used for measurements of
the sky background. We use the $3\arcsec$ diameter fibers
optimized for blue transmission, and the spectrograph is
configured with an echelle grating and an all-transmission
optics camera providing high throughput at a resolution of
$\sim$ 20,000. All observations were done at central wavelengths
of $5130$\AA\ or $6385$\AA\ with a wavelength range of $\sim 200$\AA\,
providing rich arrays of narrow absorption lines. Radial velocities
with a precision of $\la 0.5~km s^{-1}$ are derived from the spectra
via cross-correlation with a high $S/N$ sky spectrum \citep{hmm+06,mbd+01}.

\begin{figure}[ht!]
\epsscale{1.0}
\plotone{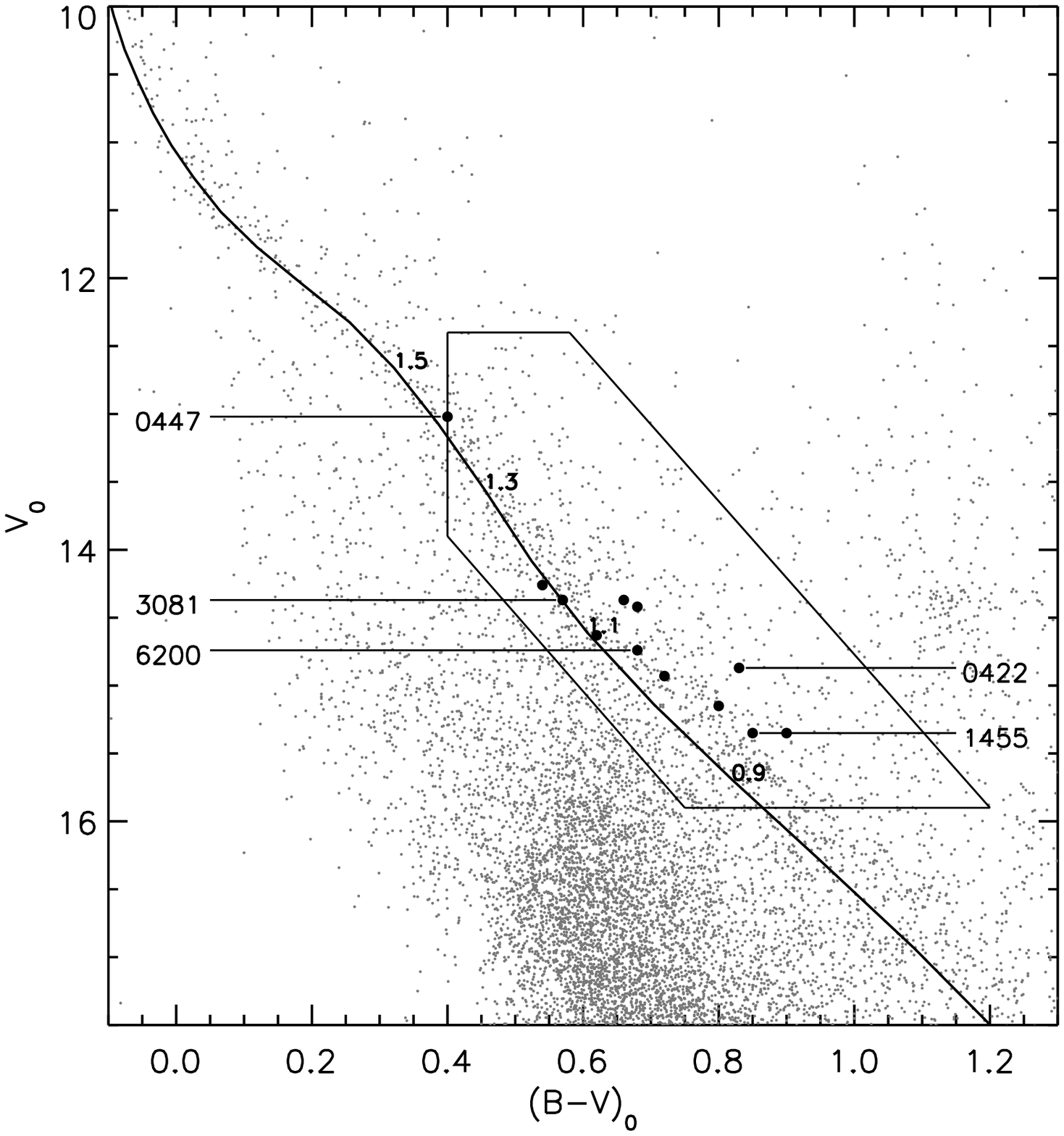}
\caption{The color-magnitude diagram of M35. The photometry
and the cluster reddening ($E_{(B-V)} = 0.2$) were provided by
\citet{deliyannis06}. The 12 binaries with known rotation periods
of their primary stars are marked as black dots. The five binaries
with orbital period shortward of 13 days are labeled with the
last four digits of their respective 2MASS ID's. The 150 Myr isochrone
over-plotted has been corrected for reddening and extinction and
a distance modulus of $9\fm8$ \citep{kfr+03}. Relevant isochrone masses
are marked. The stars included in our spectroscopic survey fall
within the region outlined.
\label{cmd}}
\end{figure}

The initial selection of target stars was based on photometric
cluster membership in the color-magnitude diagrams (see
Figure~\ref{cmd}). For M35 proper-motion membership studies
to $V \lesssim 15$ by \citet{ms86} and \citet{cudworth71}
were used as well. The target list for M35
includes stars of type mid F to mid K, corresponding to a range
in stellar mass from $\sim 1.4~M_{\odot}$ ($V_{0} \simeq 12.5$,
$(B-V)_{0} \simeq$ 0.4) to $\sim 0.7~M_{\odot}$ ($V_{0} \simeq 16$,
$(B-V)_{0} \simeq$ 1.1), with solar-mass stars at $V_{0} \sim 15$.
In M34 stars of type early F to early M were observed corresponding
to a range in stellar mass from $\sim 1.5~M_{\odot}$ ($V_{0} \simeq
12.0$, $(B-V)_{0} \sim$ 0.3) to $\sim 0.4~M_{\odot}$ ($V_{0} \simeq
16.5$, $(B-V)_{0} \sim$ 1.5), with solar mass stars at $V_{0}
\sim 13.5$.

Telescope time granted from Wisconsin and NOAO\footnote{NOAO is the national
center for ground-based nighttime astronomy in the United States
and is operated by the Association of Universities for Research
in Astronomy (AURA), Inc. under cooperative agreement with the
National Science Foundation.} allowed for 3-4 spectroscopic observing
runs per year per cluster, with each run typically including
multiple observations on several sequential nights. Once identified,
velocity variables are observed at a frequency appropriate to the
timescale of their variation. At present the radial-velocity survey
of M35 has resulted in a sample of 50 spectroscopic binaries for
which orbital solutions have been derived. The orbital periods
span 2.25 days to 3112 days corresponding to separations from
0.04 to $\sim$ 5 AU, assuming a $1~M_{\odot}$ primary star and
a $0.5~M_{\odot}$ secondary star. In M34 orbital parameters have
been derived for 20 spectroscopic binaries spanning orbital periods
from 2.26 to 1210 days. This paper is based on a subset of those
binaries that are described in detail below. 

\subsection{Time-Series Photometry}

\begin{figure}[ht!]
\epsscale{1.0}
\plotone{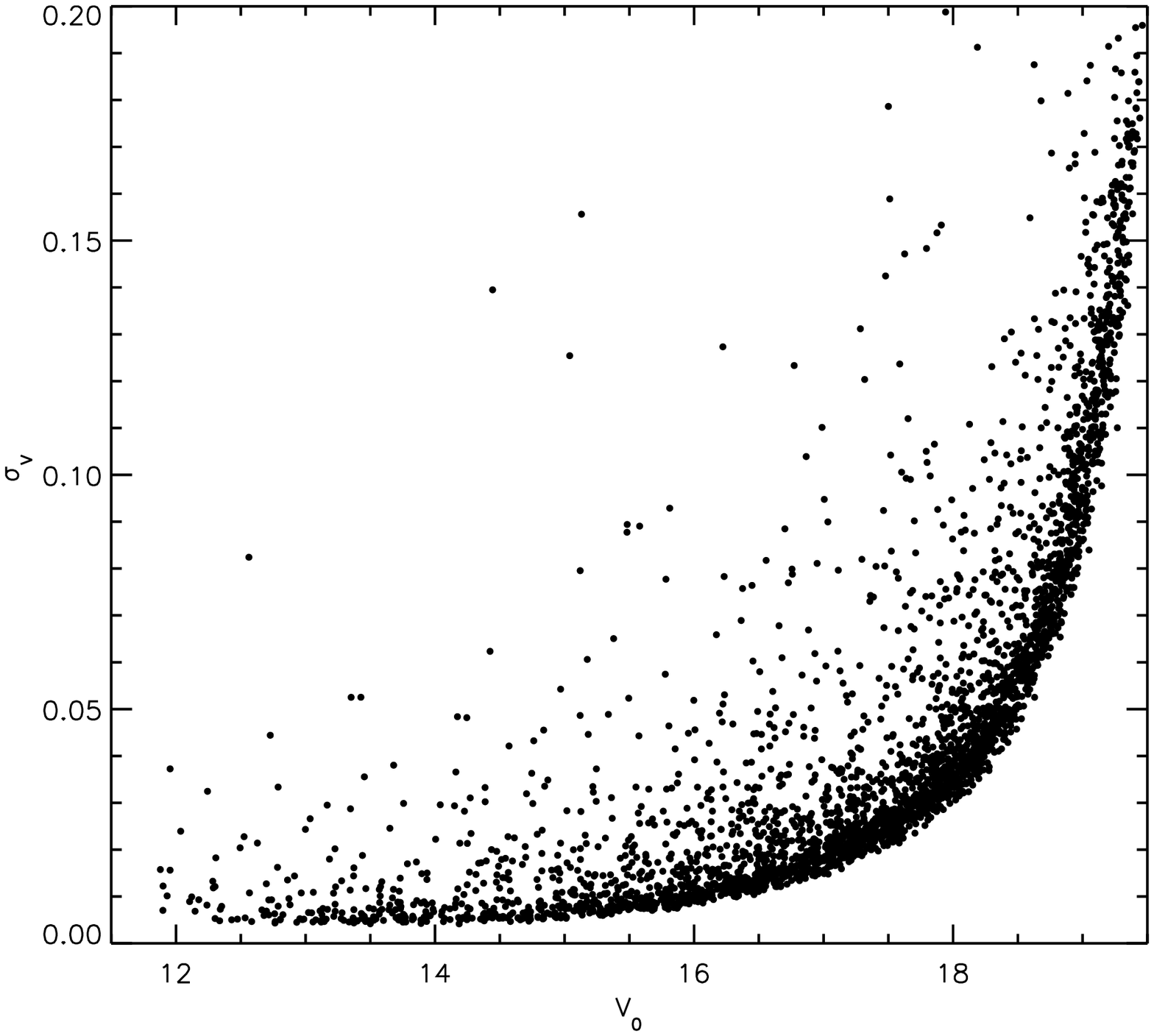}
\caption{The standard deviation of the instrumental $V$ magnitudes
as a function of the true V magnitude ($V_0$) for stars in the field
of M35. A first-order estimate of the true V magnitude has been
obtained by applying a correction of $-3\fm238$, equivalent to
the mean difference between the instrumental V magnitude and the
V magnitude from \citet{deliyannis06}, plus an extinction correction
of $3 \times E_{(B-V)} = 0\fm6$ \citep{deliyannis06}. A relative
photometric precision of $\sim$ 0.5\% is obtained for stars with
$12\fm0 \la V \la 15\fm0$.
\label{m0_sigst}}
\end{figure}

We have photometrically surveyed stars in a $\sim 40\arcmin \times 
40\arcmin$ region centered on M35 and M34. The photometric data were
obtained using the WIYN 0.9m telescope \footnote{The 0.9m telescope
is operated by WIYN Inc. on behalf of a Consortium of ten partner
Universities and Organizations (see http://www.noao.edu/0.9m/general.html)}
at Kitt Peak equipped with a $2k \times 2k$ CCD camera. The complete
dataset is composed of images from two different but complementary
observing programs. Images of the two clusters were acquired by the
first author from December $1^{st} - 17^{th}$ 2002 with a frequency
of approximately once per hour for $\sim$ six hours per night that
the clusters airmasses were below 1.5. In addition, one image per
night was obtained in a queue-scheduled observing program from
October 2002 to March 2003. We reduced
our CCD frames using the standard IRAF CCDRED package. We used
the IRAF GASP package to compute a simple linear transformation
of pixel coordinates to equatorial coordinates for each frame,
using as reference approximately 30 stars from the Digitized
Sky Survey per frame. Our derived stellar positions show a
frame-to-frame scatter of less than $0\farcs1$ in each direction.
We identified stellar sources using the IRAF DAOFIND task and
performed PSF photometry using the DAOPHOT package. The procedures
used were described and developed in \citet{smm+99} and \citet{svm+02}.
Figure~\ref{m0_sigst}
displays the standard deviation as a function of V magnitude
for stars in the field of M35. A relative photometric precision
of $\sim$ 0.5\% is obtained for stars with $12 \la V \la 15$,
with slightly poorer precision at the $V = 16\fm5$ faint-limit
of the spectroscopic study. For M35 the result of the photometric
survey is a database of differential photometric V-band light
curves for $\sim$ 14000 stars with $12 \la V_{0} \la 19.5$. At
the present time only the photometric data on M35 have been
reduced and analyzed. The photometry for star 6211 in M34
presented in this paper is kindly provided by \citet{barnes05}.

We employed the \citet{scargle82} periodogram analysis
to detect periodic variability in the light curves \citep[see][]{smm+99}.
For each candidate star, we generate a set of 100 synthetic
light curves, each consisting of normally distributed noise
with a nightly and a night-to-night dispersion representative
of our data. A periodogram was computed for each test light
curve and the maximum of the 100 observed power levels was
adopted as the level of 1\% false-alarm probability (FAP). 
This measured FAP was used as the criterion for accepting or
rejecting detected photometric variability; we accept only
periods whose maximum periodogram signals are stronger than
the power corresponding to the 1\% FAP level. From our database
we have determined stellar rotation periods for 443 stars.
Of these, 259 have one or more radial-velocity measurements
(the remainder being below the faint limit of the spectroscopic
survey or photometric non-members), 203 are photometric and
spectroscopic members of M35, and 12 are members of binary
systems with known orbital parameters. \citet{mms06b} present
and describe in more detail our photometric data, the reduction
thereof, and the methods used for detecting periodic variability.

\section{OBSERVATIONAL RESULTS} \label{results}

We present the spectroscopic and photometric results for 12 binaries
in M35 and one binary in M34. These binaries are single-lined spectroscopic
systems with well-determined orbital periods ($\sigma_{P_{orb}}/P_{orb}
\la 0.01$) and eccentricities ($\sigma_{e} \la 10^{-2}$).
The rotation periods of their primary stars are determined from
periodic variations in their light curves, presumably due to spots
on the stellar surfaces. The mass of the primary star in each binary
has been estimated from 150 Myr and 250 Myr Yale isochrones \citep{ykd03}
fitted to the cluster sequences of M35 and M34, respectively. Photometry
and values for cluster reddening for both clusters were provided by
\citet{deliyannis06} and \citet{steinhauer06}. Table 1 lists all
observational results for the 13 binaries together with the derived
orbital angular velocities and stellar rotation angular velocities
needed for studying tidal synchronization. Figure~\ref{cmd} shows the
color-magnitude diagram (CMD) of M35. The locations of these 12 binaries
are marked as black dots. Five M35 binaries discussed in detail below
are labeled with the last four digits of their respective 2MASS ID's
(see Table 1). The 150 Myr isochrone is corrected for reddening and
extinction and a distance modulus of $9\fm8$ \citep{kfr+03}. Relevant
model masses are marked along the isochrone.

\begin{deluxetable}{rcccrccrccccrrc}
\tabletypesize{\scriptsize}
\rotate
\tablecaption{Photometric and Spectroscopic Observational Results}
\tablewidth{0pt}
\tablehead{
\colhead{ID}\tablenotemark{a} &
\colhead{$V_{o}$} &
\colhead{$(B-V)_{o}$} &
\colhead{$\gamma$} &
\colhead{$P_{orbit}$} &
\colhead{$e$} &
\colhead{$\sigma_{e}$} &
\colhead{$P^{prim}_{rot}$} &
\colhead{$\delta(P^{prim}_{rot})$}\tablenotemark{b} &
\colhead{$M_{prim}$} &
\colhead{$\omega$}\tablenotemark{c} &
\colhead{$\Omega_{\star}$}\tablenotemark{d} &
\colhead{$\Omega_{\star}/\Omega_{ps}$}\tablenotemark{e} &
\colhead{$\log(\Omega_{\star}/\Omega_{ps})$} &
\colhead{$P_{RV}$}\tablenotemark{f} \\
\colhead{} &
\colhead{} &
\colhead{} &
\colhead{($km~s^{-1}$)} &
\colhead{(Days)} &
\colhead{} &
\colhead{} &
\colhead{(Days)} &
\colhead{(Days)} &
\colhead{($M_{\odot}$)} &
\colhead{($rad~day^{-1}$)} &
\colhead{($rad~day^{-1}$)} &
\colhead{} &
\colhead{} &
\colhead{\%}
}
\startdata
 06090257+2420447 & 13.016 & 0.403 & -8.30 &   10.28 & 0.009 & 0.019 &  2.30 & 0.02  & 1.4 & 0.611 & 2.73 &   4.46 &  0.65 & 94 \\ 
 06090306+2420095 & 14.369 & 0.655 & -6.92 & 3112.67 & 0.394 & 0.118 &  2.48 & 0.01  & 1.1 & 0.002 & 2.54 & 625.66 &  2.80 & 90 \\ 
 06090306+2419361 & 15.150 & 0.802 & -9.25 &  637.03 & 0.234 & 0.025 &  4.70 & 0.07  & 1.0 & 0.010 & 1.34 & 101.74 &  2.01 & 91 \\ 
 06090352+2417234 & 15.351 & 0.896 & -6.74 &  156.60 & 0.580 & 0.034 &  2.38 & 0.02  & 0.9 & 0.040 & 2.64 &  17.52 &  1.24 & 88 \\ 
 06091557+2410422 & 14.867 & 0.834 & -6.14 &    8.17 & 0.649 & 0.022 &  3.71 & 0.06  & 0.9 & 0.769 & 1.69 &   0.44 & -0.35 & 76 \\ 
 06091924+2417223 & 14.933 & 0.724 & -8.86 &  795.30 & 0.255 & 0.056 &  5.25 & 0.08  & 1.0 & 0.008 & 1.20 & 108.47 &  2.04 & 93 \\ 
 06092436+2426200 & 14.741 & 0.678 & -7.36 &   10.33 & 0.016 & 0.009 & 10.13 & 0.39  & 1.1 & 0.609 & 0.62 &   1.02 &  0.01 & 93 \\ 
 06095563+2417454 & 14.420 & 0.680 & -7.54 &   30.13 & 0.273 & 0.005 &  2.84 & 0.03  & 1.1 & 0.209 & 2.22 &   7.29 &  0.86 & 94 \\ 
 06085441+2403081 & 14.368 & 0.565 & -7.45 &   12.28 & 0.550 & 0.003 &  6.03 & 0.12  & 1.1 & 0.512 & 1.04 &   0.61 & -0.21 & 93 \\ 
 06082017+2421514 & 14.259 & 0.544 & -8.16 & 2324.11 & 0.199 & 0.093 &  2.56 & 0.02  & 1.2 & 0.003 & 2.45 & 731.43 &  2.86 & 94 \\ 
 06074436+2430262 & 14.634 & 0.624 & -7.04 &  476.21 & 0.389 & 0.046 &  4.26 & 0.07  & 1.1 & 0.013 & 1.48 &  56.47 &  1.75 & 91 \\ 
 06083789+2431455 & 15.354 & 0.855 & -8.08 &    2.25 & 0.010 & 0.008 &  2.29 & 0.02  & 0.9 & 2.794 & 2.74 &   0.98 & -0.01 & 93 \\ 
~\\
02410619+4246211\tablenotemark{g} & 15.323 & 1.000 & -7.27 &    4.39 & 0.063 & 0.033 &  8.03 & 0.10 & 0.7 & 1.431 & 0.78 & 0.53 & -0.28 & 94 \\
\enddata

\tablenotetext{a}{Stellar 2MASS ID.}
\tablenotetext{b}{The estimated uncertainty of the stellar rotation period ($\delta(P^{prim}_{rot}$)).}
\tablenotetext{c}{Average orbital angular velocity.}
\tablenotetext{d}{Measured rotational angular velocity of the primary star.}
\tablenotetext{e}{Ratio of the measured rotational angular velocity to the expected pseudosynchronous rotational angular velocity.}
\tablenotetext{f}{The radial velocity membership probability ($P_{RV}$) calculated using the formalism by \citet[see text]{vkp58}.}
\tablenotetext{g}{M34 binary.}

\end{deluxetable}

\begin{figure}[ht!]
\epsscale{1.0}
\plottwo{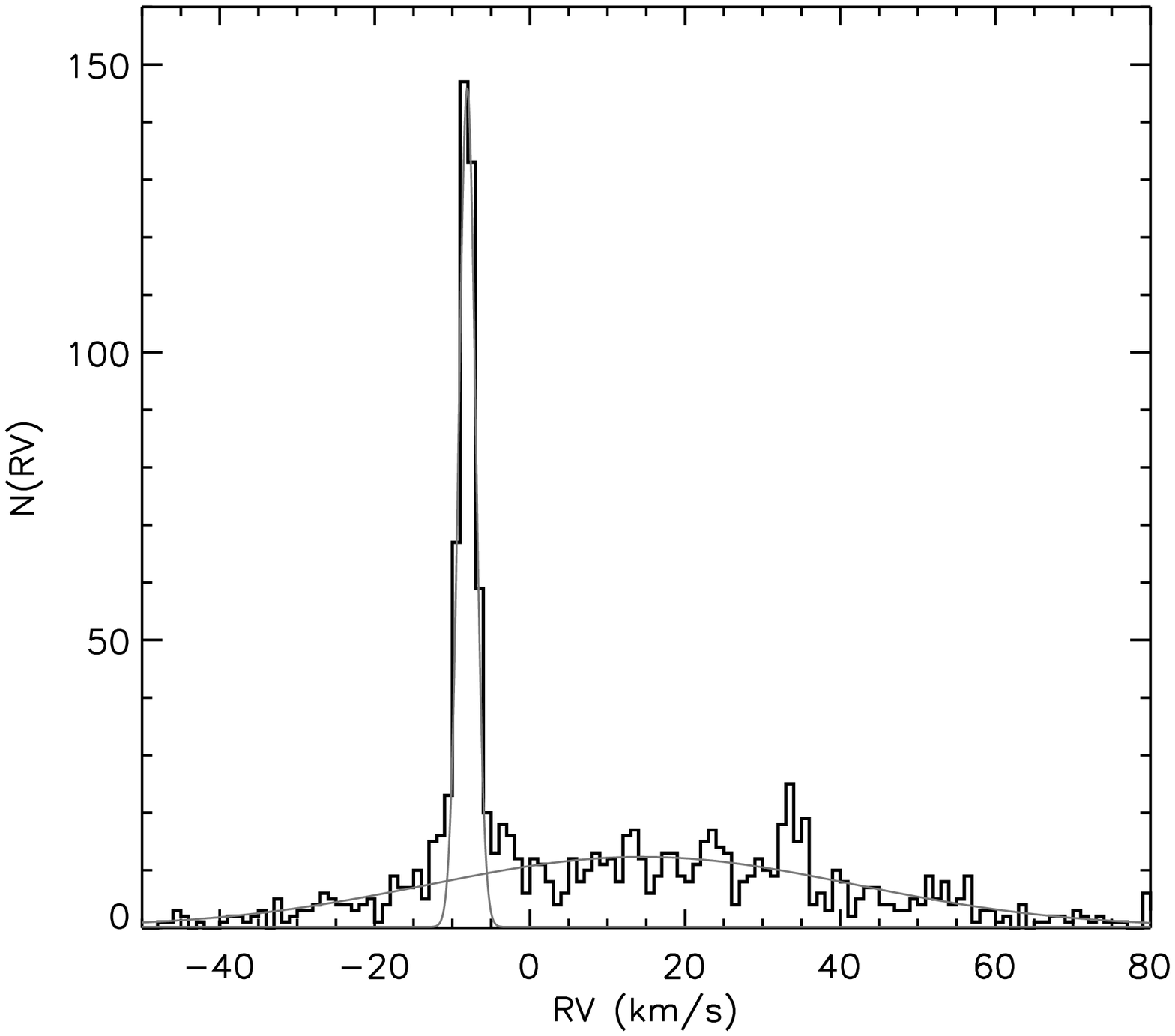}{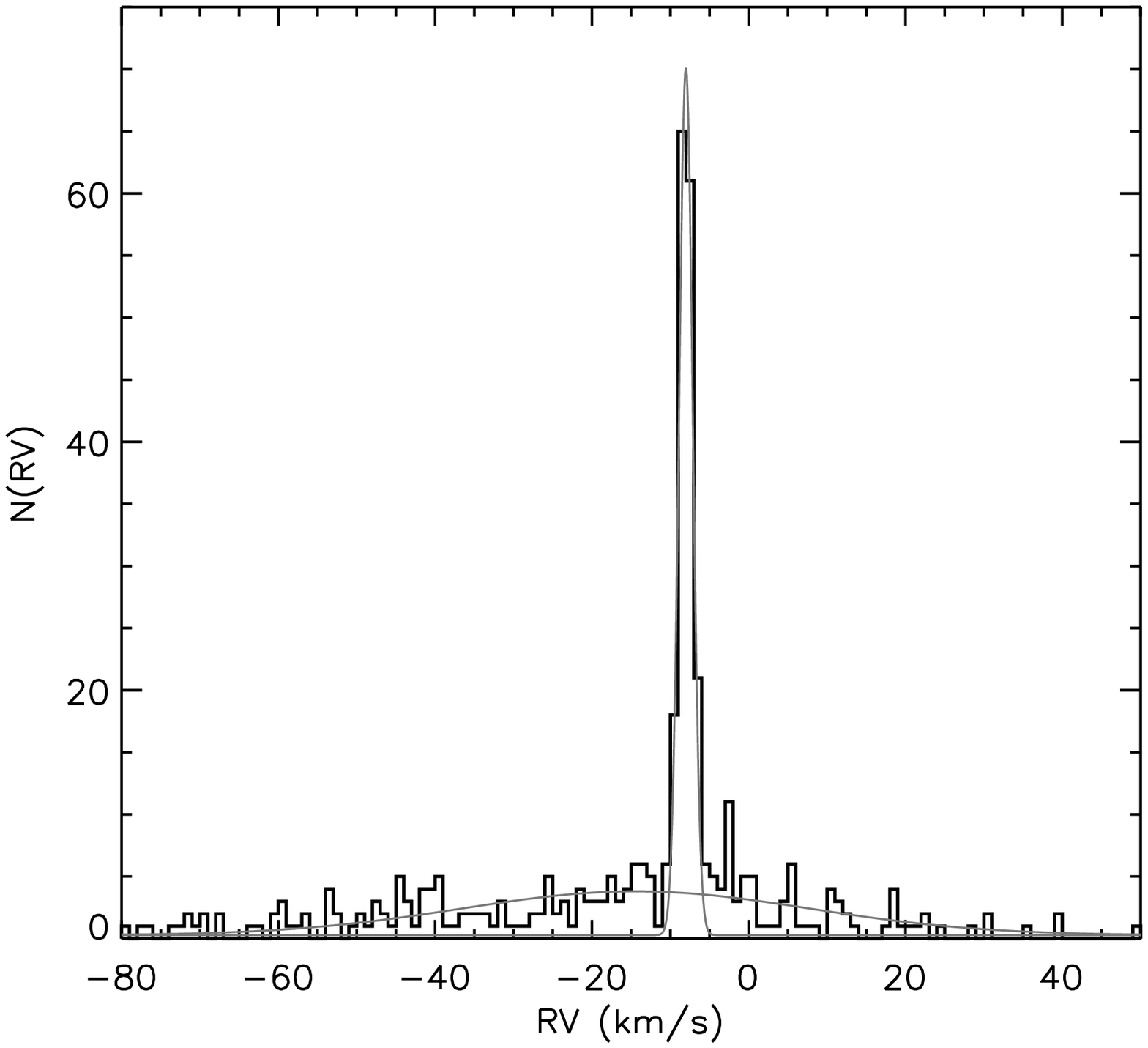}
\caption{The distributions of radial-velocities for stars
in the fields of M35 (left) and M34 (right). Two Gaussian
functions (grey solid curves) have been simultaneously
fitted to the cluster and field components of each distribution.
\label{vel}}
\end{figure}

All 13 binaries are photometric and radial-velocity members
of M35 or M34. Figure~\ref{vel} shows the distributions of
measured radial-velocities for the two clusters. Gaussian
functions have been simultaneously fitted to the cluster
and field components of each distribution. The radial-velocity
cluster membership probability ($P_{RV}$) of each of the
13 binary stars is calculated following the formalism by \citet{vkp58}

\begin{equation}
P_{RV} = \frac{C(RV)}{C(RV) + F(RV)}
\end{equation}

\noindent where $C(RV)$ and $F(RV)$ represent the values of
the Gaussian fit to the cluster and field distributions,
respectively, for the center-of-mass radial-velocity of a binary.

\begin{figure}[ht!]
\epsscale{1.0}
\plottwo{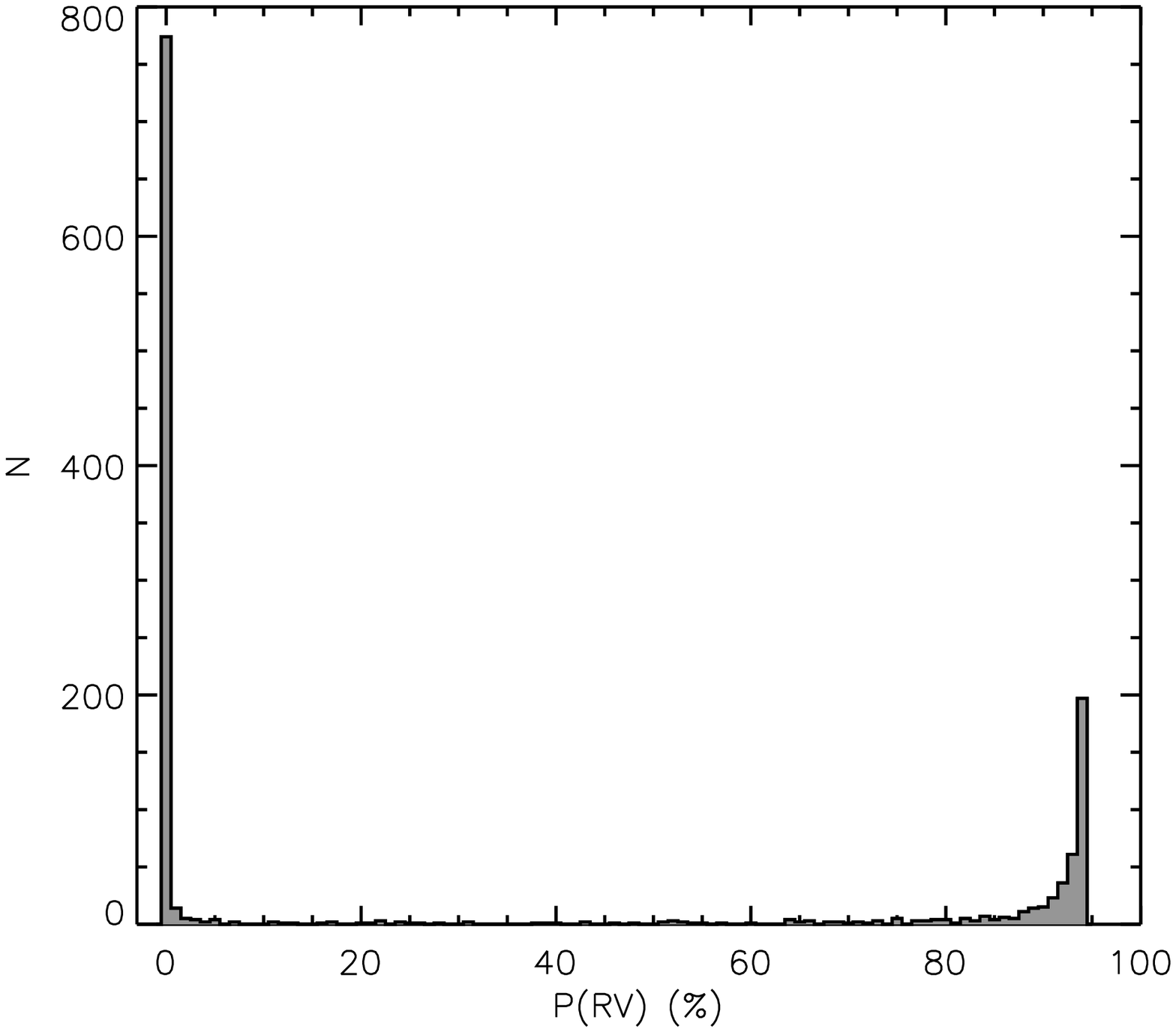}{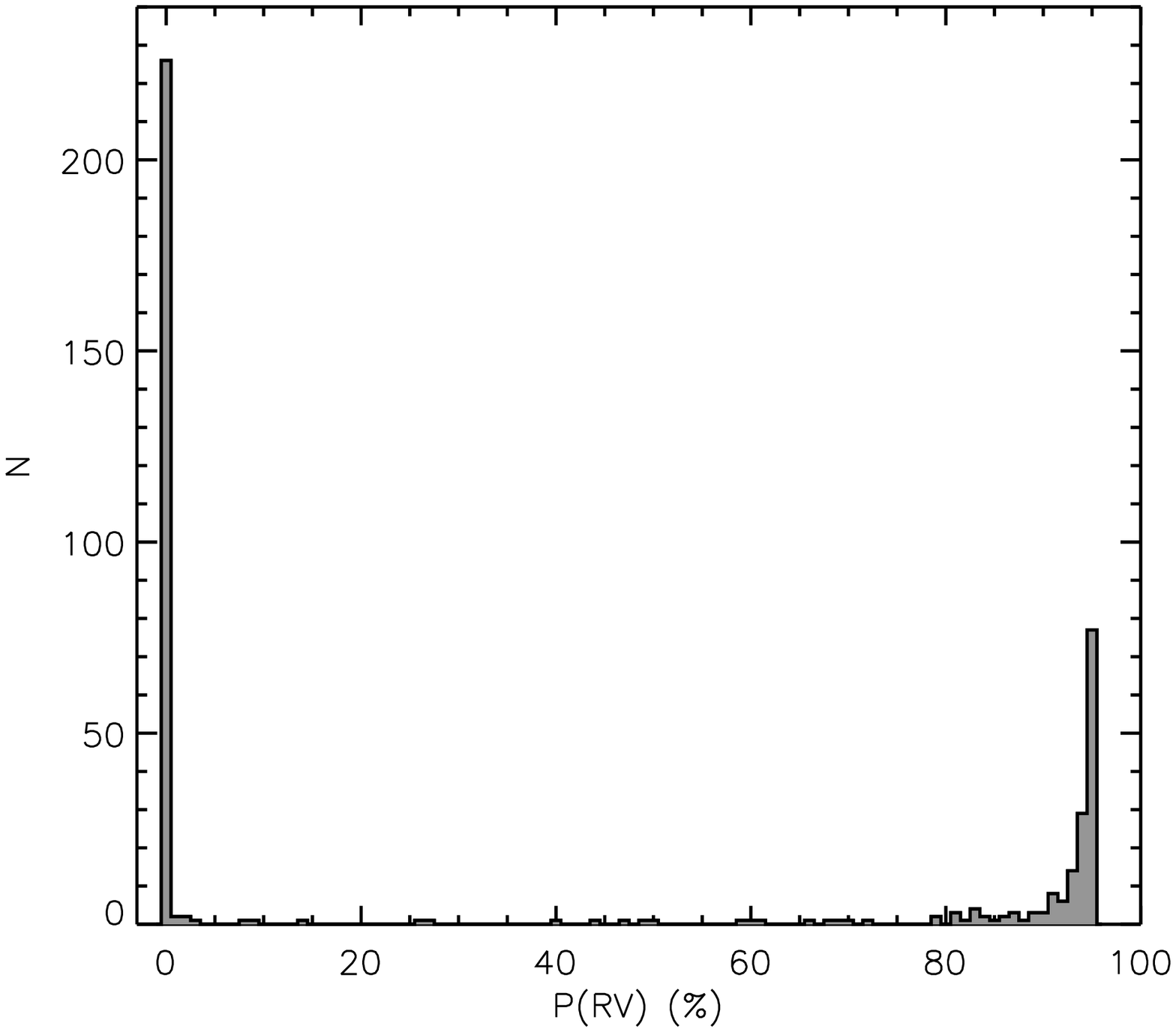}
\caption{The distributions of radial-velocity membership
probabilities (P(RV)) for stars in the fields of M35 (left)
and M34 (right). Cluster members and non-members are easily
identified as two distinct peaks in the distributions.
The space between the peaks is populated by stars with
radial-velocities corresponding to the wings of the cluster
distributions.
\label{memprobs}}
\end{figure}

Figure~\ref{memprobs} shows the distributions of radial-velocity
membership probabilities for stars in the fields of M35 and M34.
In both clusters the separation between members and non-members
is distinct. Less than 10\% of the stars have probabilities
placing them between the member and non-member peaks, corresponding
to radial-velocities on the wings of the cluster distributions.

Of particular interest to our study of tidal synchronization
are the six binaries with orbital periods in the range from 2.25
days to 12.28 days. The stars in these six systems are close
enough that their spins and orbits have evolved due to tidal
interactions. Of the six, the orbital parameters for the 5
M35 binaries were first presented in \citet{mm05}. The photometric
light curves are presented here and the derived rotation periods
are listed in Table 1. The uncertainty ($\pm \delta P$) on the
rotation periods were determined using the expression for the
periodogram resolution \citep{kovacs81}

\begin{equation}
\delta P = P^2 \frac{3 \sigma}{4 T \sqrt{N} A}
\end{equation}

\noindent where $\sigma$ is the uncertainty in the photometric
data, T is the total time spanned by the data, N is the number
of independent data points, and A is the amplitude of the detected
signal. When estimating rotation period uncertainties we made the
conservative assumption that only data from separate nights are
truly independent and set the value of N to the number of nights
of data in the light curve. We describe here in detail the
observational results for those six systems.

\begin{figure}[ht!]
\epsscale{1.0}
\plotone{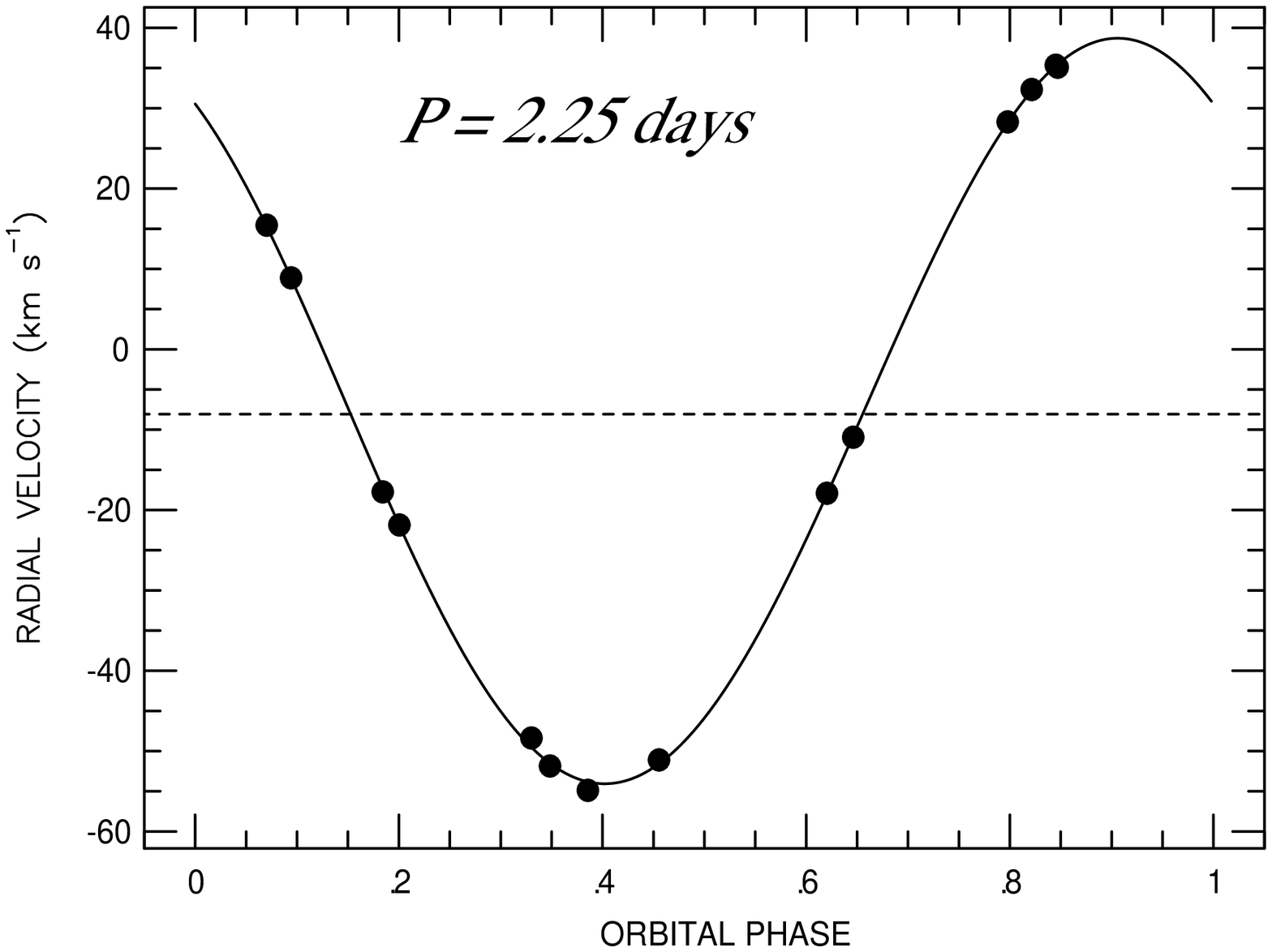}
\caption{The radial velocity measurements of binary 1455 phased
to an orbital period of 2.25 days. The best fit orbital solution
is over-plotted. The orbit is very near circular with an eccentricity
of only $0.010 \pm 0.003$.
\label{1455orb}}
\end{figure}

{\it Binary 1455:}
14 radial-velocity measurements have been obtained of this
binary over $\sim$ 600 orbital cycles. With an orbital period of
only 2.25 days this is the shortest period binary found in our
survey of M35. The orbit is very near circular with $e = 0.010 \pm 0.003$.
Figure~\ref{1455orb} shows the orbital solution over-plotted
on the radial-velocity data phased to the 2.25 day period. The CMD
location of binary 1455 is on the cluster main sequence slightly
above the position on the 150 Myr isochrone corresponding to
a mass of $\sim 0.9~M_{\odot}$. We use $0.9~M_{\odot}$ as an
estimate for the mass of the primary star as there is no sign
of the secondary star in the spectra/cross-correlation function
of this binary. The radial-velocity cluster membership probability
($P_{RV}$, eq. [5]) of binary 1455 is 94\%.

\begin{figure}[ht!]
\epsscale{1.0}
\plottwo{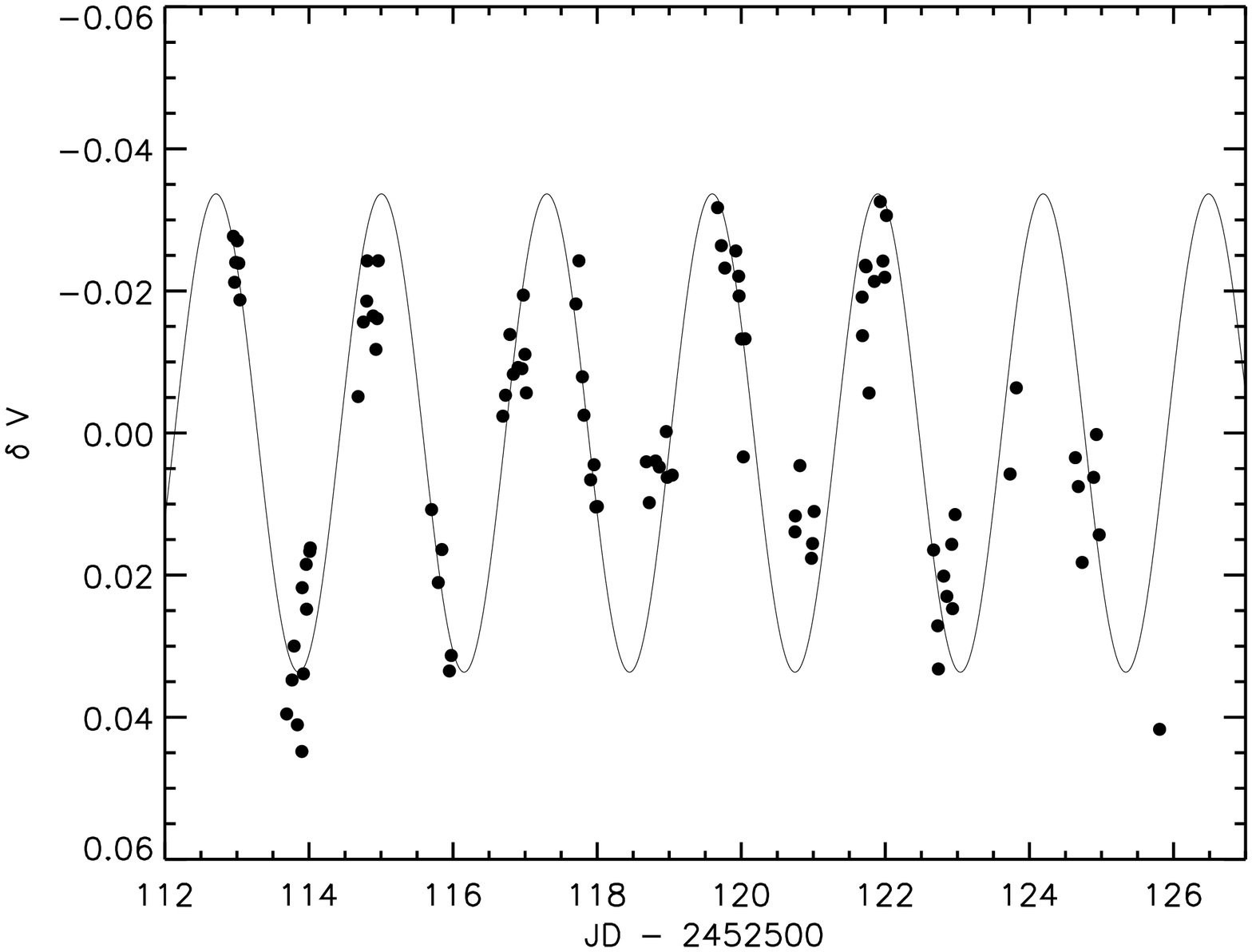}{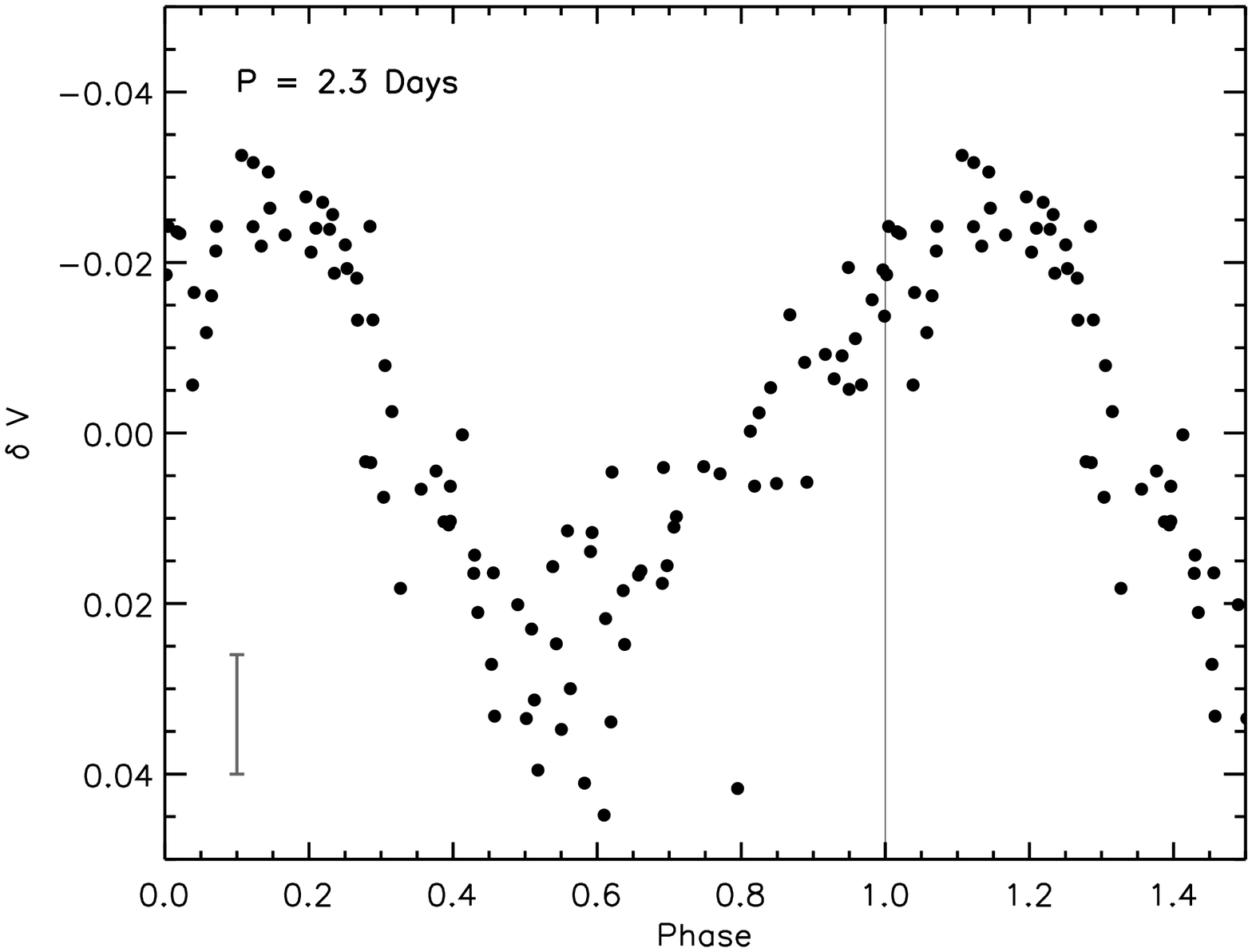}
\caption{{\bf Left:} The light curve for binary 1455 based on 96
photometric measurements from 15 nights in December 2002. A 
sine function with the a 2.29 days period overplotted.
{\bf Right:} The differential V-band
photometry for binary 1455 phased to a period of $2.29 \pm 0.02$ days,
corresponding to the maximum periodogram power. The vertical solid
line indicates a phase value of 1.0, and the errorbar in the lower
left-hand corner represent $\pm$ the typical photometric error at
the V magnitude of binary 1455.
\label{1455lc}}
\end{figure}

The light curve and phased light curve shown in Figure~\ref{1455lc}
are based on 96 photometric measurements obtained over 15 nights
in December 2002. The maximum periodogram power corresponds
to a period of $2.29 \pm 0.02$ days. We note that phasing the
photometric measurements with the binary orbital period of 2.25
days rather than the independently determined value of 2.29 days
produces only a subtle change of the light curve. When including
the queue data a total of 156 measurements were obtained
from October 2002 to March 2003. The same rotation period
was found using all 156 measurements, but the phased light curve
is noisier, presumably due to the varying quality of the queue
data and possibly due to irregularities in the spot modulation
over the longer timescale.

\begin{figure}[ht!]
\epsscale{1.0}
\plotone{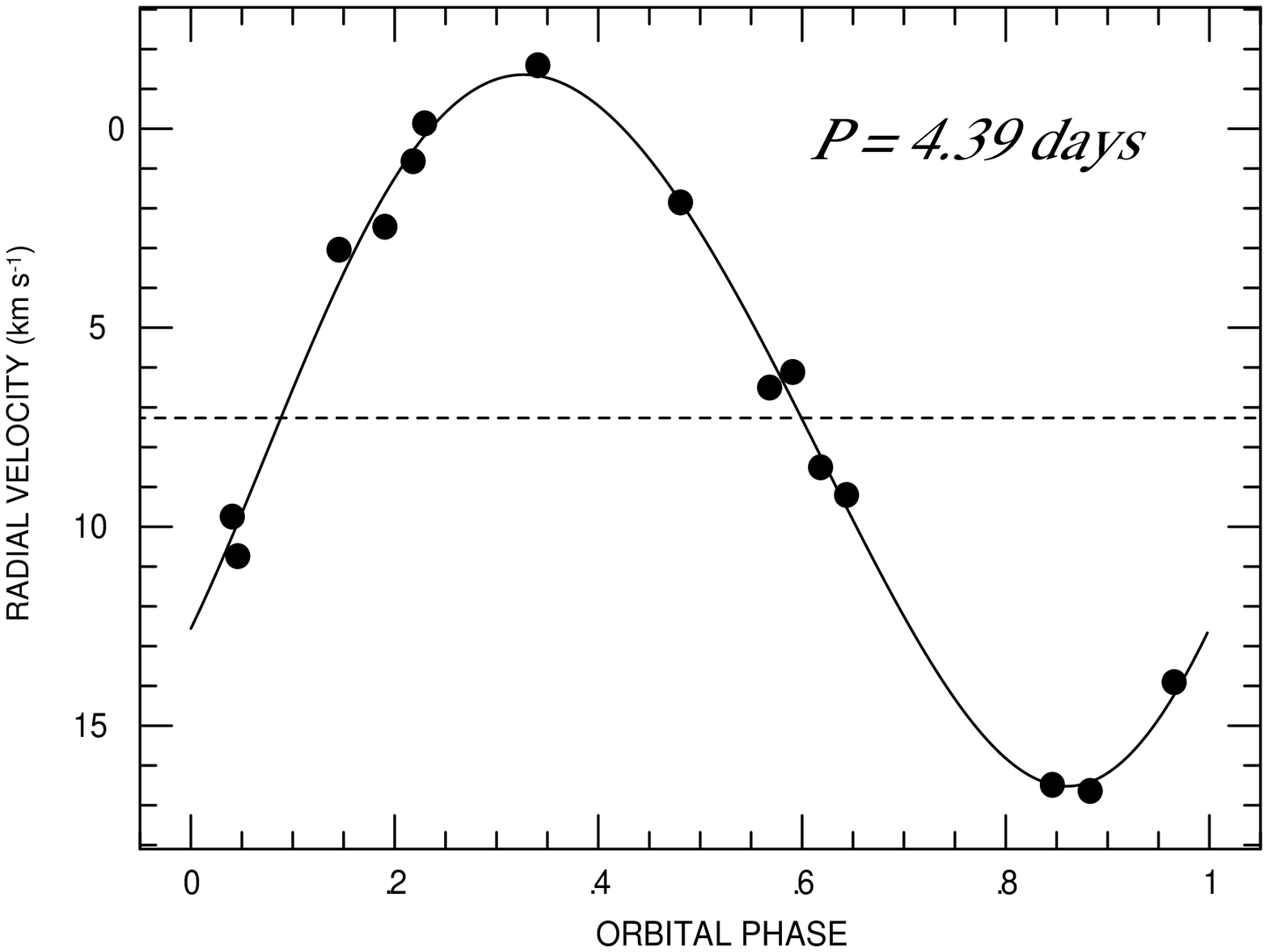}
\caption{The radial velocity measurements of binary 6211
phased to an orbital period of 4.39 days. The best fit orbital
solution is over-plotted. The orbit is circular with an
eccentricity of $0.063 \pm 0.033$.
\label{6211orb}}
\end{figure}

{\it Binary 6211:}
Figure~\ref{6211orb}
shows the orbital solution over-plotted
on 15 radial-velocity measurements over $\sim$ 160 orbital cycles.
The data have been phased to a 4.39 day period. The orbit is
indistinguishable from circular with an eccentricity of
$0.063 \pm 0.033$. We estimate the mass of the primary star to be
$\sim 0.7~M_{\odot}$ using the fit of a 250 Myr Yale isochrone
to the M34 cluster sequence. The radial-velocity cluster
membership probability of binary 6211 is 94\%.

\begin{figure}[ht!]
\epsscale{1.0}
\plottwo{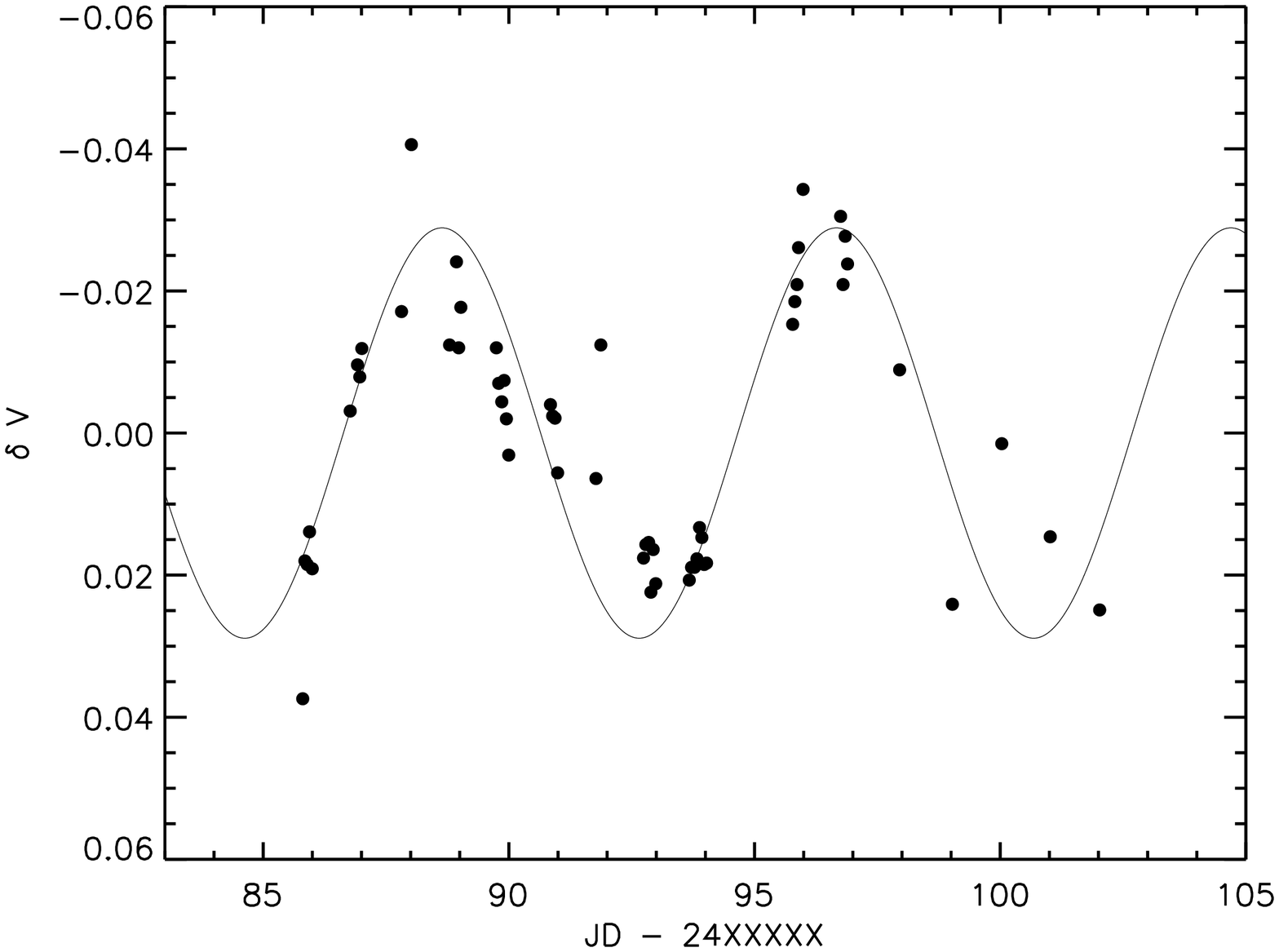}{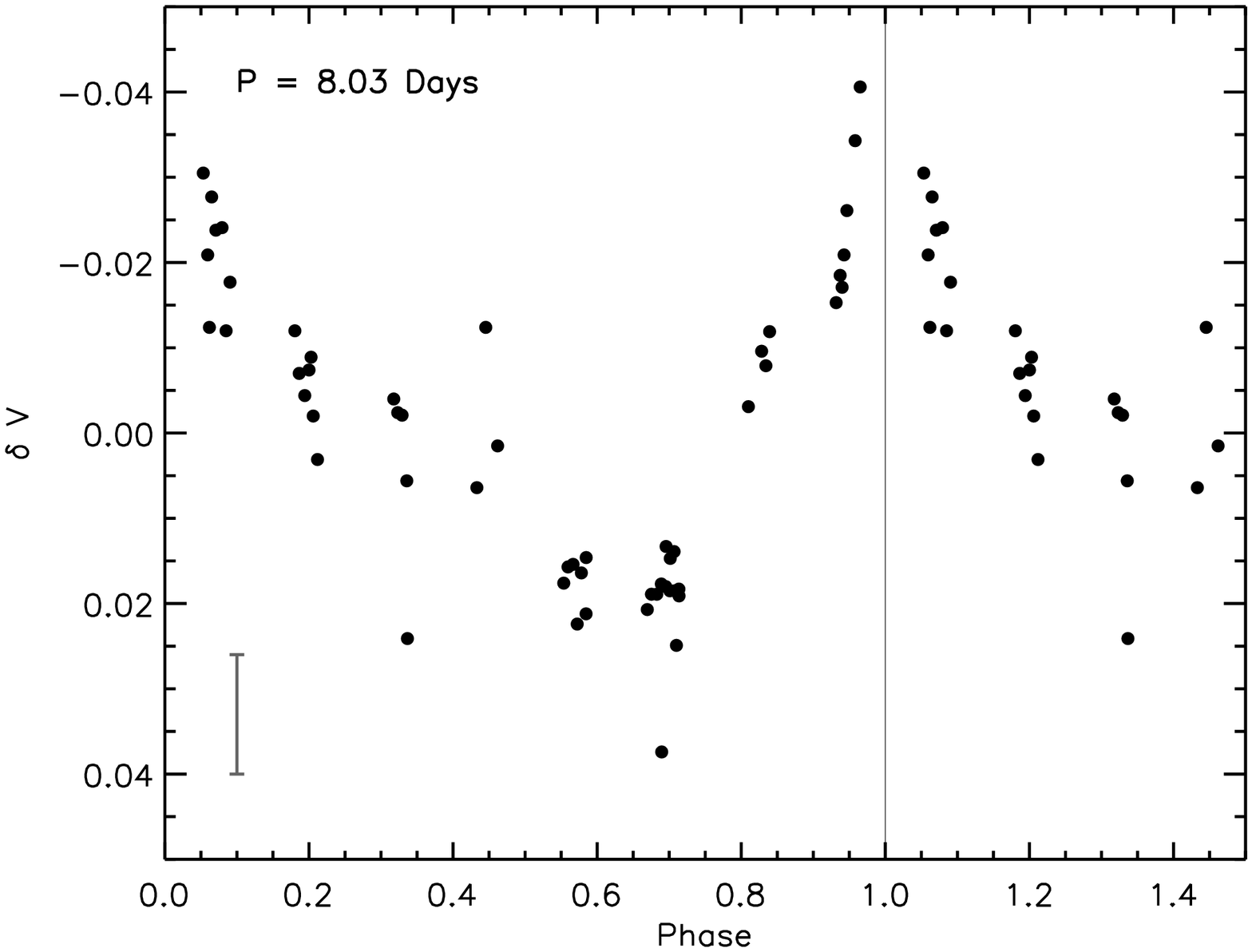}
\caption{{\bf Left:} The light curve for binary 6211 based
on 55 photometric measurements kindly provided by \citet{barnes05}.
A sine function with the a period of 8.03 days is overplotted.
{\bf Right:} The differential V-band photometry for binary 6211
\citep{barnes05} phased to a period of $8.03 \pm 0.1$ days.
The vertical solid line indicates a phase value of 1.0, and
the errorbar in the lower left-hand corner represent $\pm$
the typical photometric error at the V magnitude of binary
6211 \citep{barnes05}.
\label{6211lc}}
\end{figure}

The light curve and phased light curve shown in Figure~\ref{6211lc}
are based on 55 photometric measurements over 15 nights. The data
were kindly provided by \citet{barnes05}. The maximum power in the
periodogram corresponds to a period of $8.03 \pm 0.1$ days. We note
that phasing the photometric measurements with the binary orbital
period of 4.39 days does not lead to well-phased data.

\begin{figure}[h!]
\epsscale{1.0}
\plotone{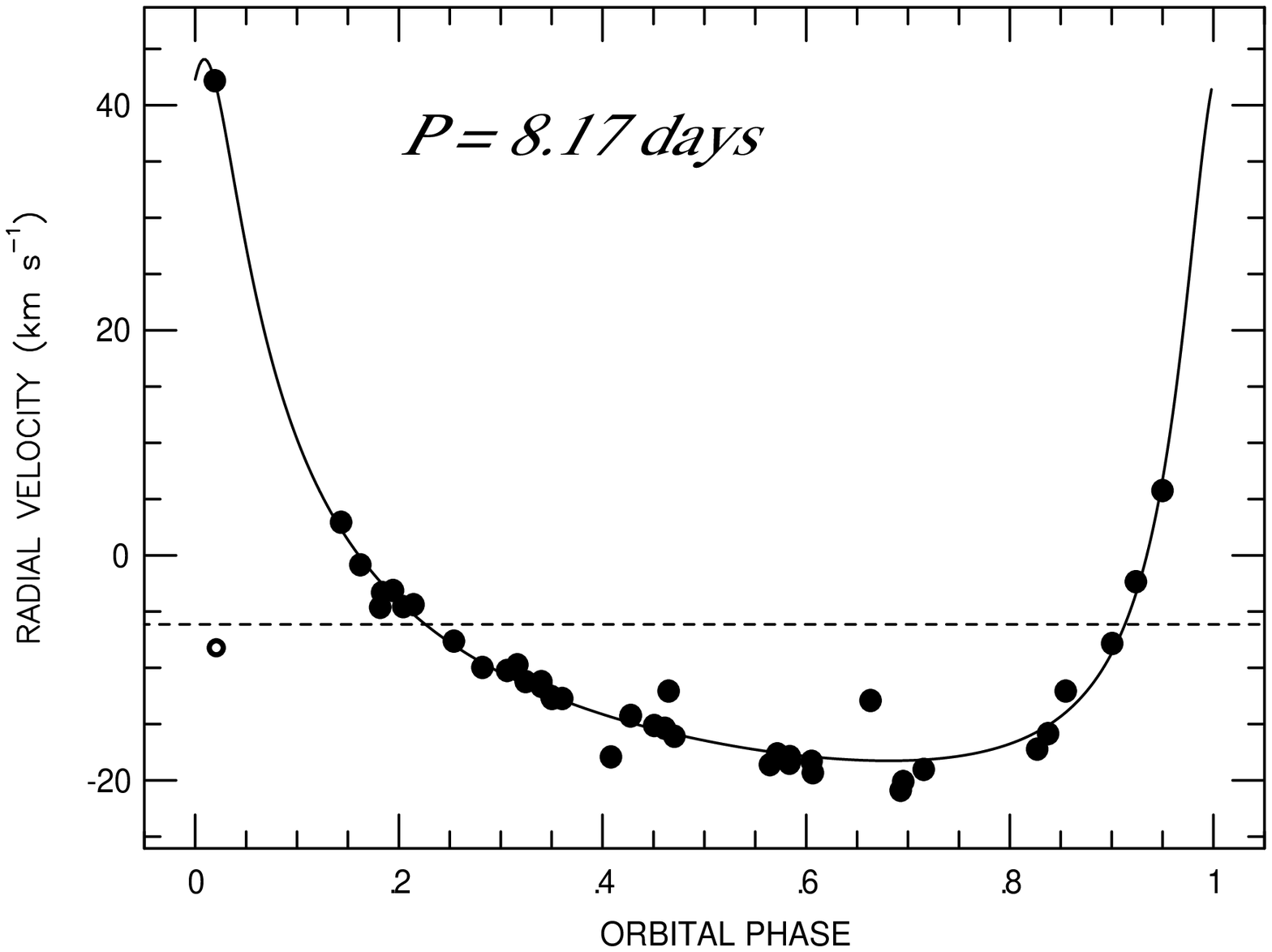}
\caption{The radial velocity measurements of
binary 0422 phased to an orbital period of 8.17 days.
The best fit orbital solution is over-plotted. The orbit
is highly eccentric ($e = 0.649 \pm 0.022)$. One observation
was obtained at periastron passage (maximum velocity separation)
revealing a third spectral component. The velocity of the tertiary
component is marked as an open circle at phase 0.04.
\label{0422orb}}
\end{figure}

{\it Binary 0422:}
The radial velocity of this binary has been measured 41 times
over $\sim$ 330 orbital cycles. The high eccentricity of the orbit
makes it difficult to observe during the short periastron passage.
One observation has been obtained close to periastron passage,
allowing a better determination of the orbital eccentricity.
The cross-correlation function from this observation revealed
a second spectral component with peak-height about two thirds
that of the primary peak and a radial-velocity of $-8.8~km~s^{-1}$,
consistent with the cluster radial velocity. We suggest therefore
that this is a triple system consisting of a close binary and a
distant tertiary star. The radial-velocity curve for the binary,
phased to a period of 8.17 days, is shown in Figure~\ref{0422orb}.
The velocity of the presumed tertiary star is marked as
a circle at phase 0.04. Over-plotted is the best fit orbital solution
with an eccentricity of $0.649 \pm 0.022$. The center-of-mass velocity
is $-6.14~km~s^{-1}$ ($\sim 2~\sigma_{cluster}$ away from the
$-8.1~km~s^{-1}$ cluster velocity), corresponding to a radial
velocity membership probability of 76\%. This deviation from the
cluster velocity may be partly due to the dynamical influence of
the triple system. The system is located $\sim 0\fm5$ above the
cluster main sequence, likely due to the combined light of the
close binary and the tertiary star. Because the tertiary star is
fainter than the primary in binary 0422, we assume that it is also
redder and we estimate the mass of the primary star in binary 0422
by assuming that in the absence of the tertiary star the binary
will be on the main-sequence fainter and bluer than the triple system.
We note however, that these assumptions have no large effect on
the estimated $1.0~M_{\odot}$ of the primary star.

\begin{figure}[h!]
\epsscale{1.0}
\plottwo{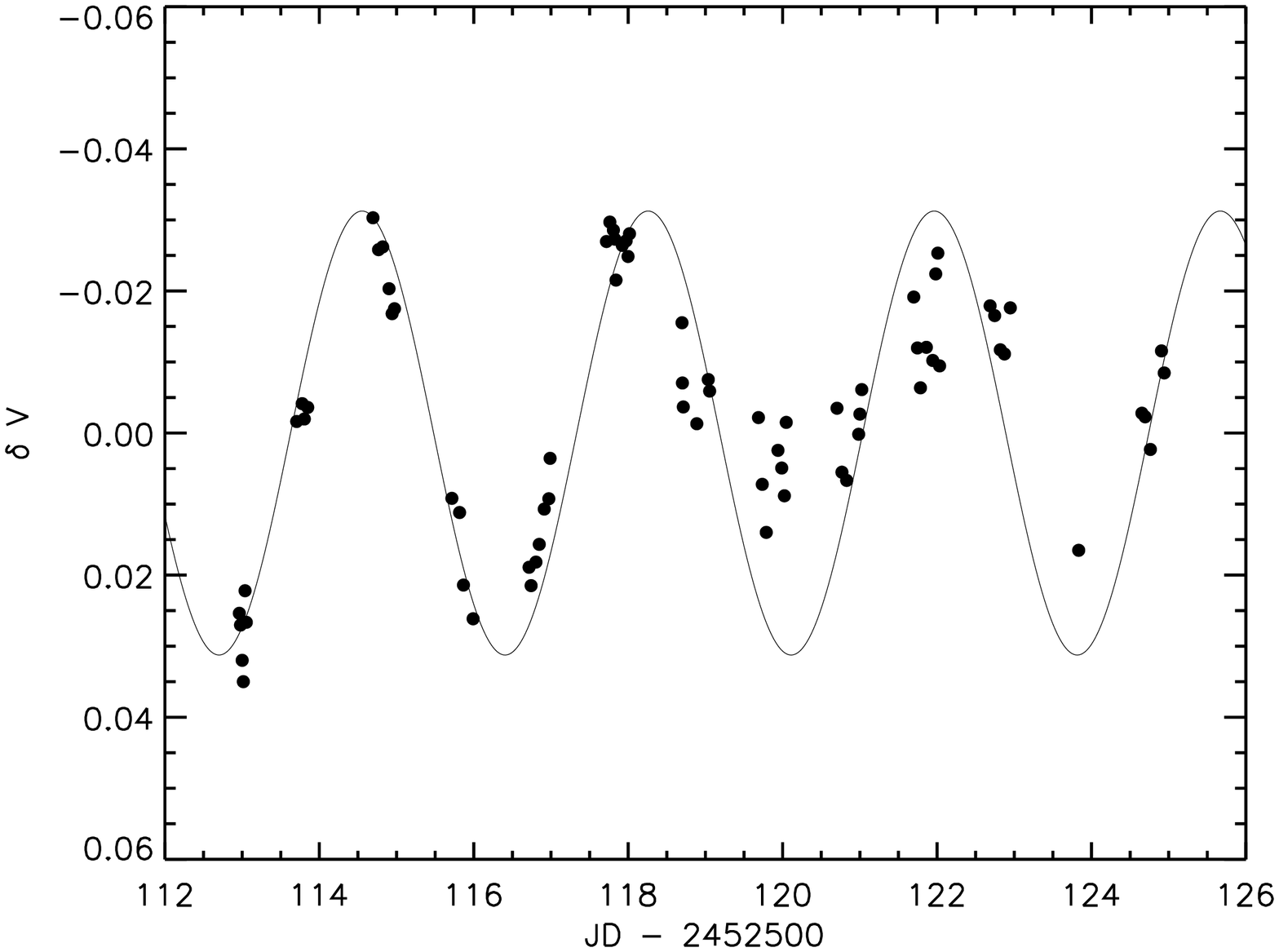}{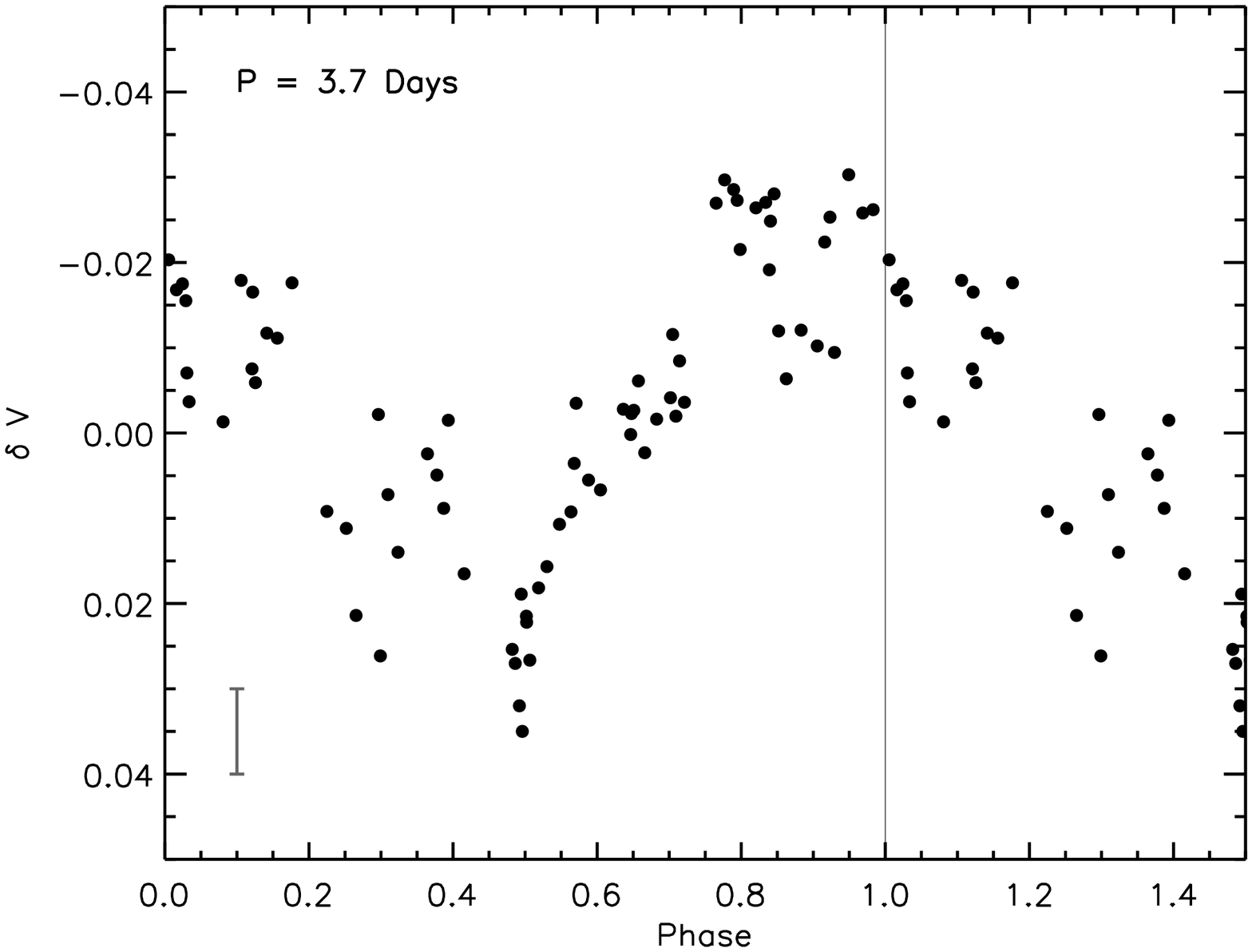}
\caption{{\bf Left:} The light curve for binary 0422 based
on 74 photometric measurements from December 2002. A sine
function with the a 3.71 days period overplotted.
{\bf Right:} The differential V-band photometry for binary 0422
phased to a period of $3.71 \pm 0.06$ days.
The vertical solid line indicates a phase value of 1.0,
and the errorbar in the lower left-hand corner represent
$\pm$ the typical photometric error at the V magnitude of binary 0422.
\label{0422lc}}
\end{figure}

Binary 0422 was imaged 126 times, 74 of which fell within 15 nights
in December 2002. The light curve and phased light curve shown in
Figure~\ref{0422lc} are based on those 74 photometric measurements.
The rotation period corresponding to the maximum periodogram power
is $3.71 \pm 0.06$ days. A slightly shorter rotation period of 3.56
days is found using all 126 photometric measurements, but the light
curve is noisier. Again, the added noise is presumably due to the
varying quality of the synoptic data and possibly due to irregularities
in the spot modulation over the longer time-scale.

\begin{figure}[h!]
\epsscale{1.0}
\plotone{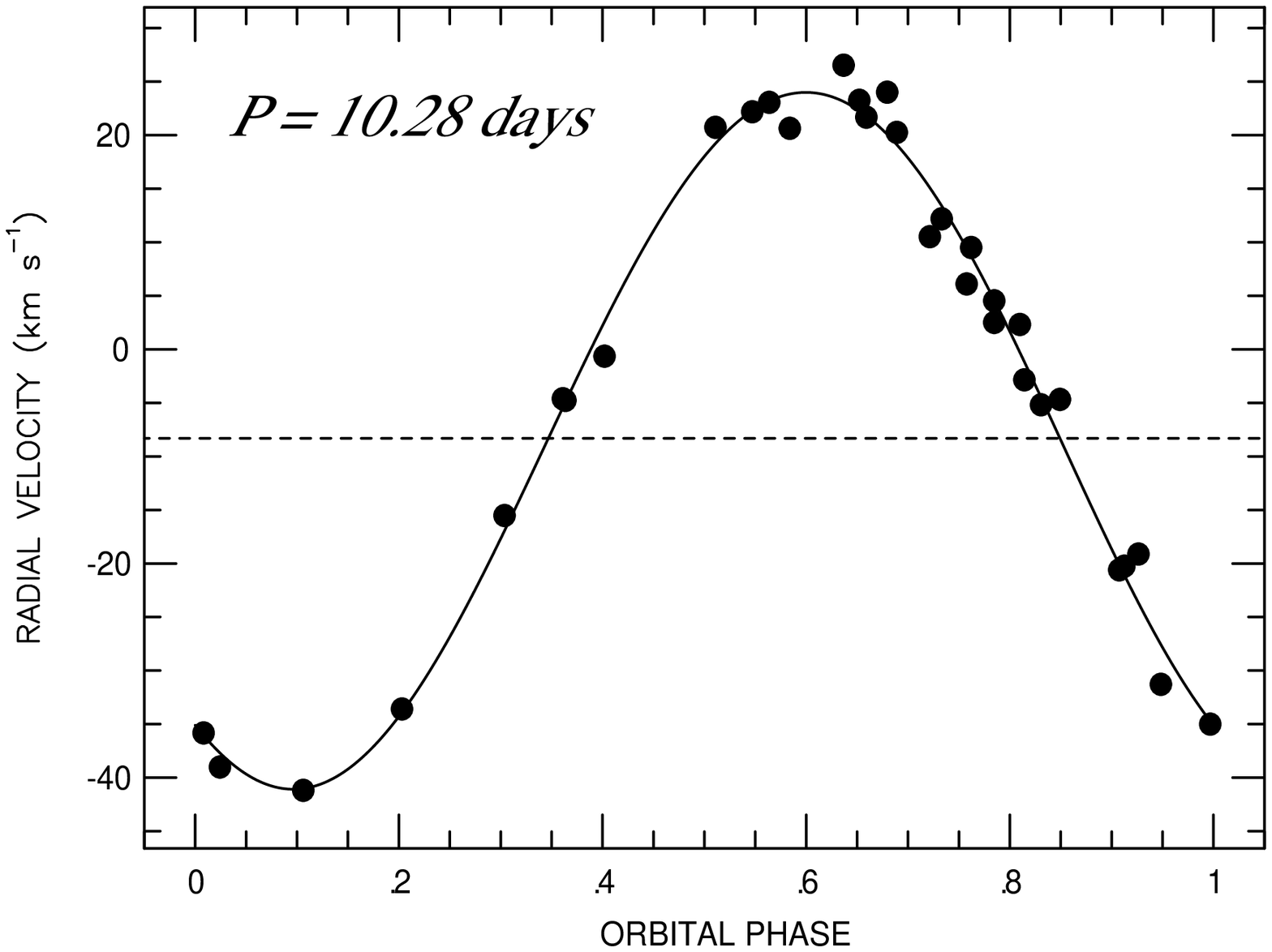}
\caption{The radial velocity measurements of
binary 0447 phased to an orbital period of 10.28 days.
The best fit orbital solution is over-plotted. The orbit
is circular ($e = 0.009 \pm 0.019$). 
\label{0447orb}}
\end{figure}

{\it Binary 0447:}
The orbital parameters of this binary were determined
from 32 radial-velocity measurements over $\sim$ 150
orbital cycles. The 10.28 day orbit is circular with
an eccentricity of $0.009 \pm 0.019$. Figure~\ref{0447orb}
shows the orbital solution over-plotted on the
phased radial-velocity data. The color and V magnitude
of binary 0447 places it on the cluster main sequence at the
blue limit of our sample of M35 stars. This
position corresponds to a mass of $\sim 1.4~M_{\odot}$ on
the 150 Myr Yale isochrone. The radial-velocity cluster
membership probability is 94\%. Furthermore,
the measured Lithium abundance for this star is consistent
with cluster membership \citep{sd04}.

\begin{figure}[h!]
\epsscale{1.0}
\plottwo{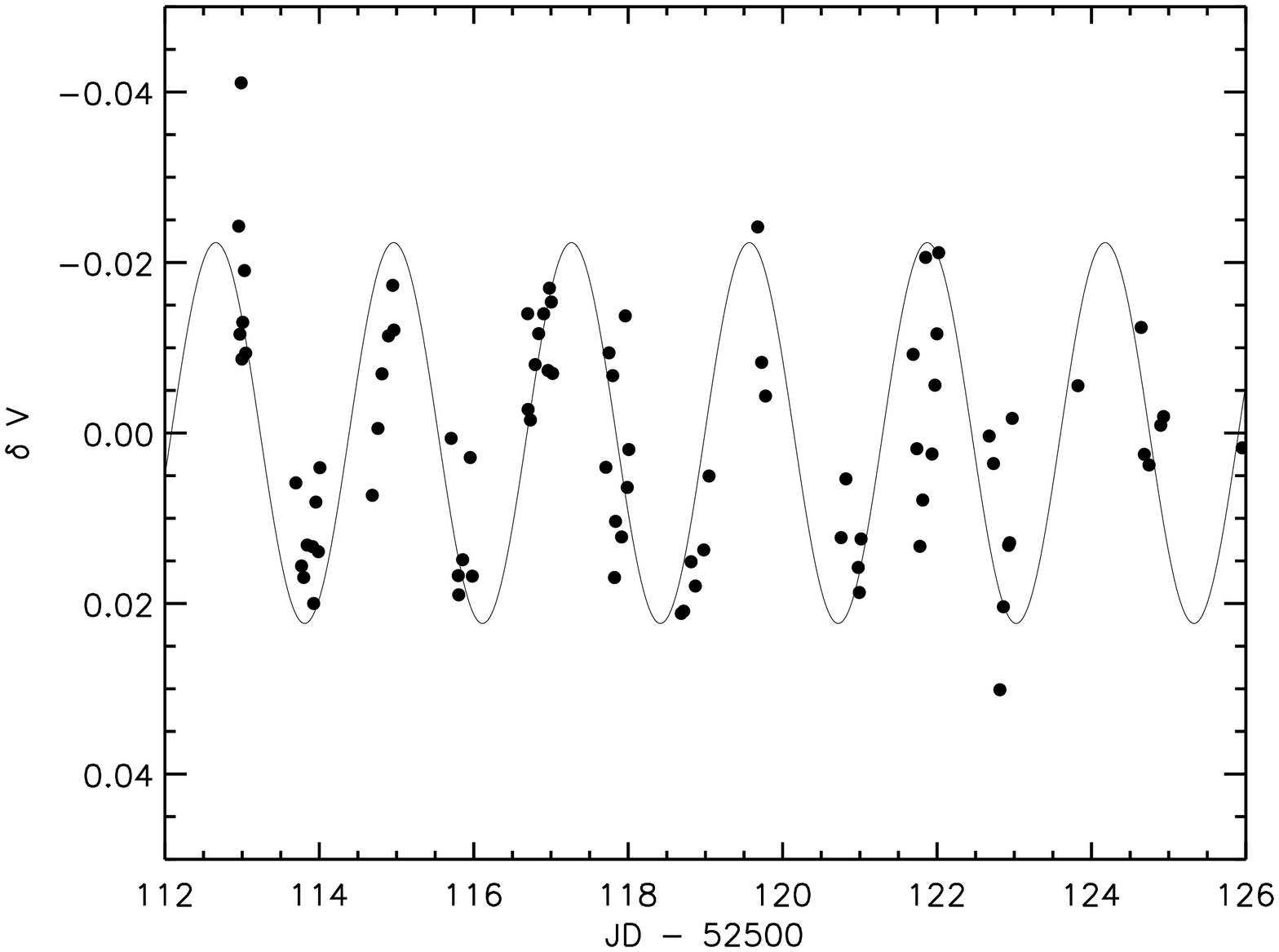}{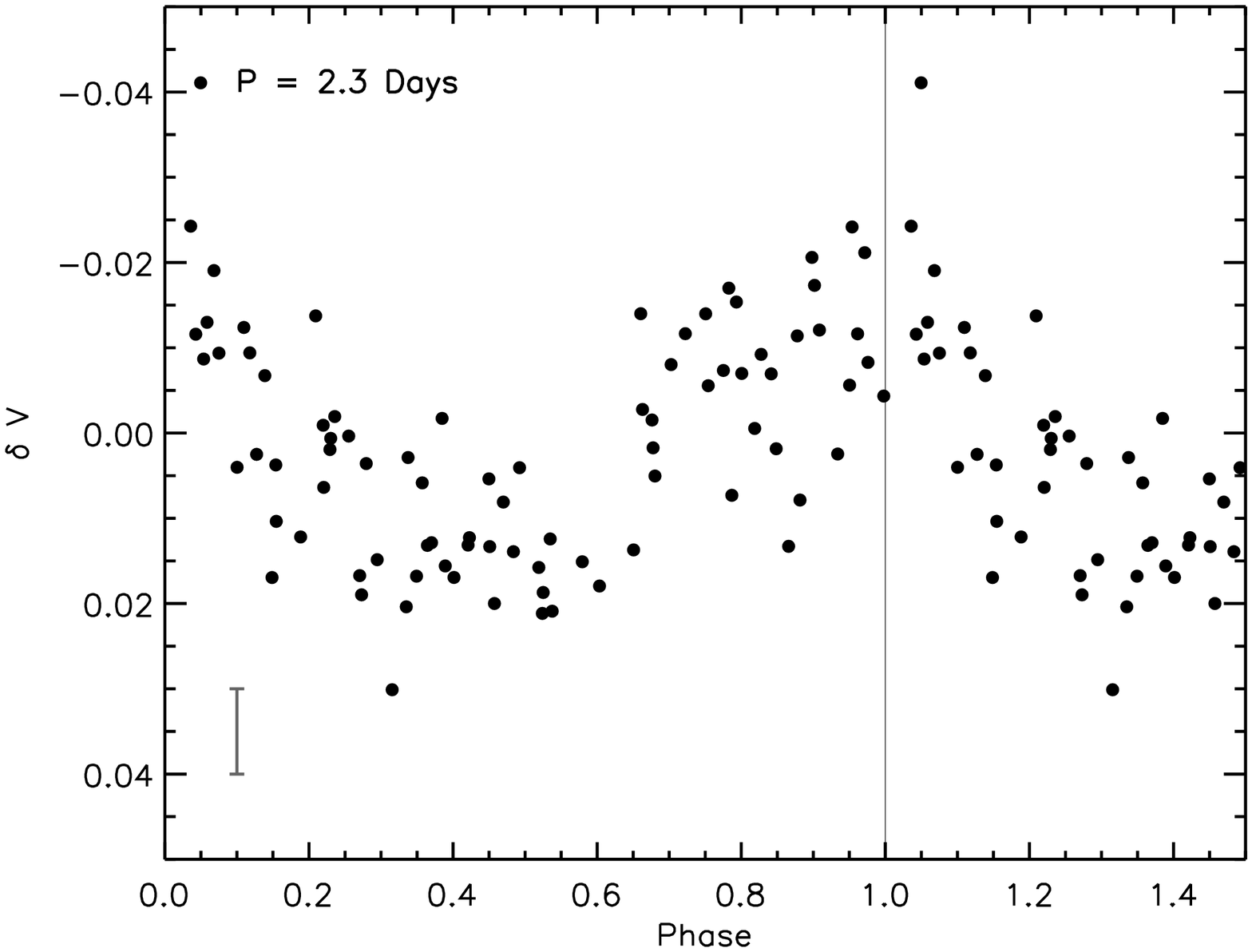}
\caption{{\bf Left:} The light curve for binary 0447 based
on 84 photometric measurements from December 2002. A sine
function with the a 2.30 days period overplotted. {\bf Right:}
The differential V-band photometry for binary 0447 phased
to a period of $2.30 \pm 0.02$ days, corresponding to
the maximum periodogram power. The vertical solid line
indicates a phase value of 1.0, and the errorbar in the
lower left-hand corner represent $\pm$ the typical photometric
error at the V magnitude of binary 0447.
\label{0447lc}}
\end{figure}

The light curve and phased light curve shown in Figure~\ref{0447lc}
are based on 84 photometric measurements from the 15 nights
in December 2002. The rotation period corresponding to the
maximum periodogram signal is $2.30 \pm 0.02$ days. A slightly
shorter rotation period of 2.2 days is found when including
the synoptic data.

\begin{figure}[h!]
\epsscale{1.0}
\plotone{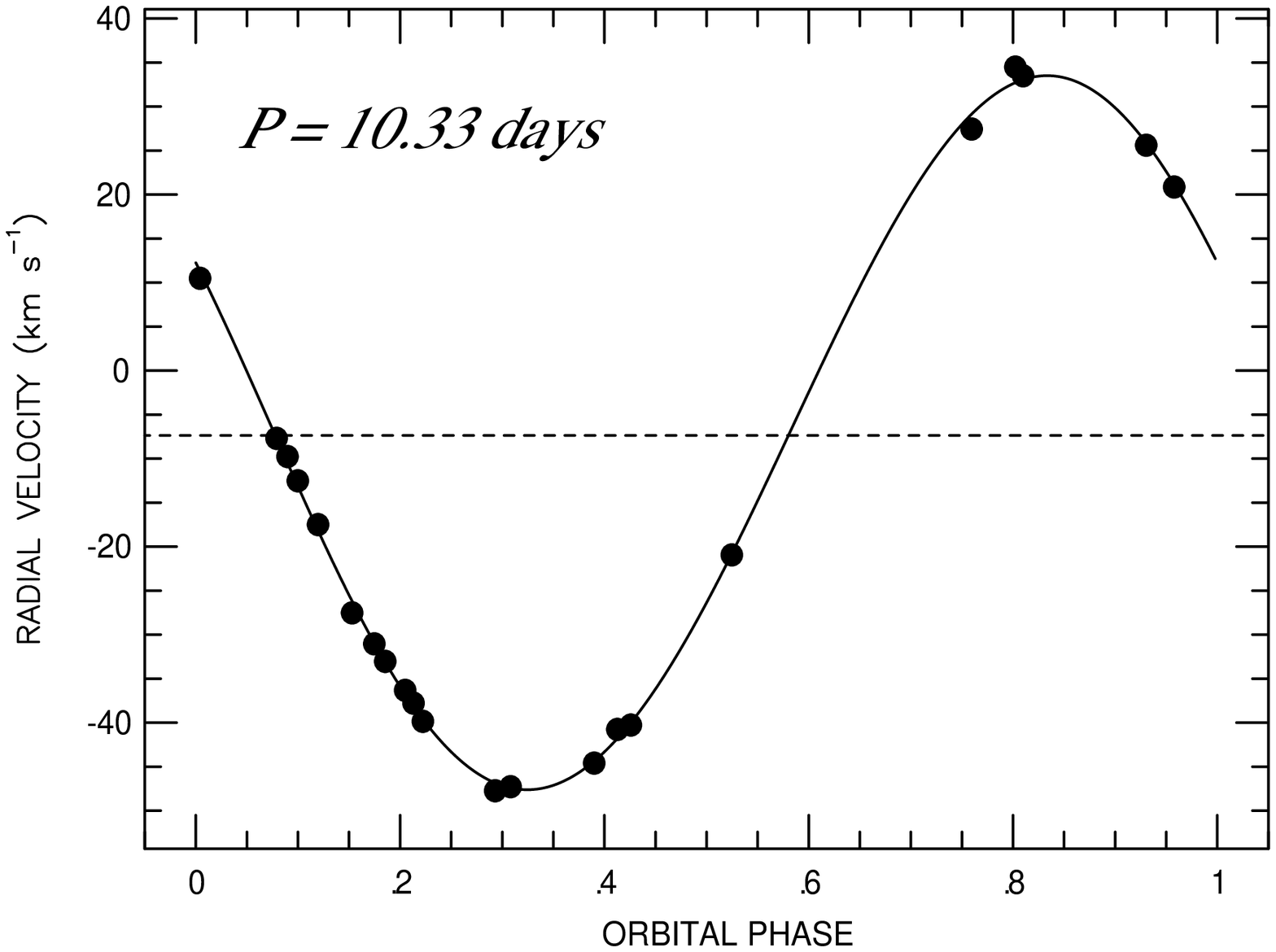}
\caption{The radial velocity measurements of
binary 6200 phased to an orbital period of 10.33 days.
The best fit orbital solution is over-plotted. The orbit
is circular ($e = 0.016 \pm 0.009$).
\label{6200orb}}
\end{figure}

{\it Binary 6200:}
Like binary 0447 this binary has a circular orbit ($e =
0.016 \pm 0.009$) with a period of $\sim$ 10 days.
Figure~\ref{6200orb} shows the orbital solution
over-plotted on the radial velocity data phased to the 10.33
day period. The orbital parameters are determined
from 22 radial velocities over $\sim$ 120 orbital cycles.
The location of binary 6200 on the M35 cluster sequence
corresponds to a mass of $\sim 1.1~M_{\odot}$.
The radial-velocity cluster membership probability is 93\%.

\begin{figure}[h!]
\epsscale{1.0}
\plottwo{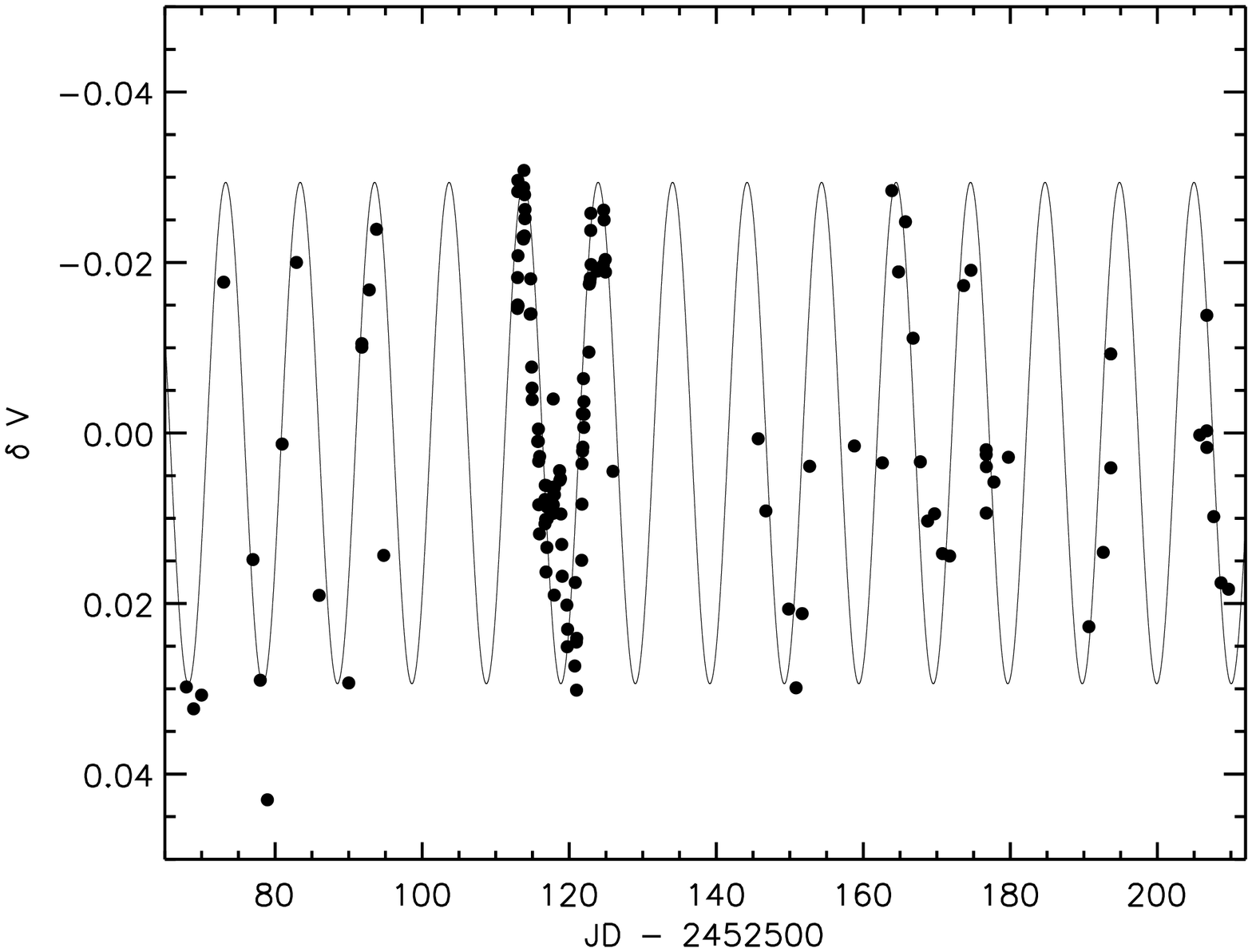}{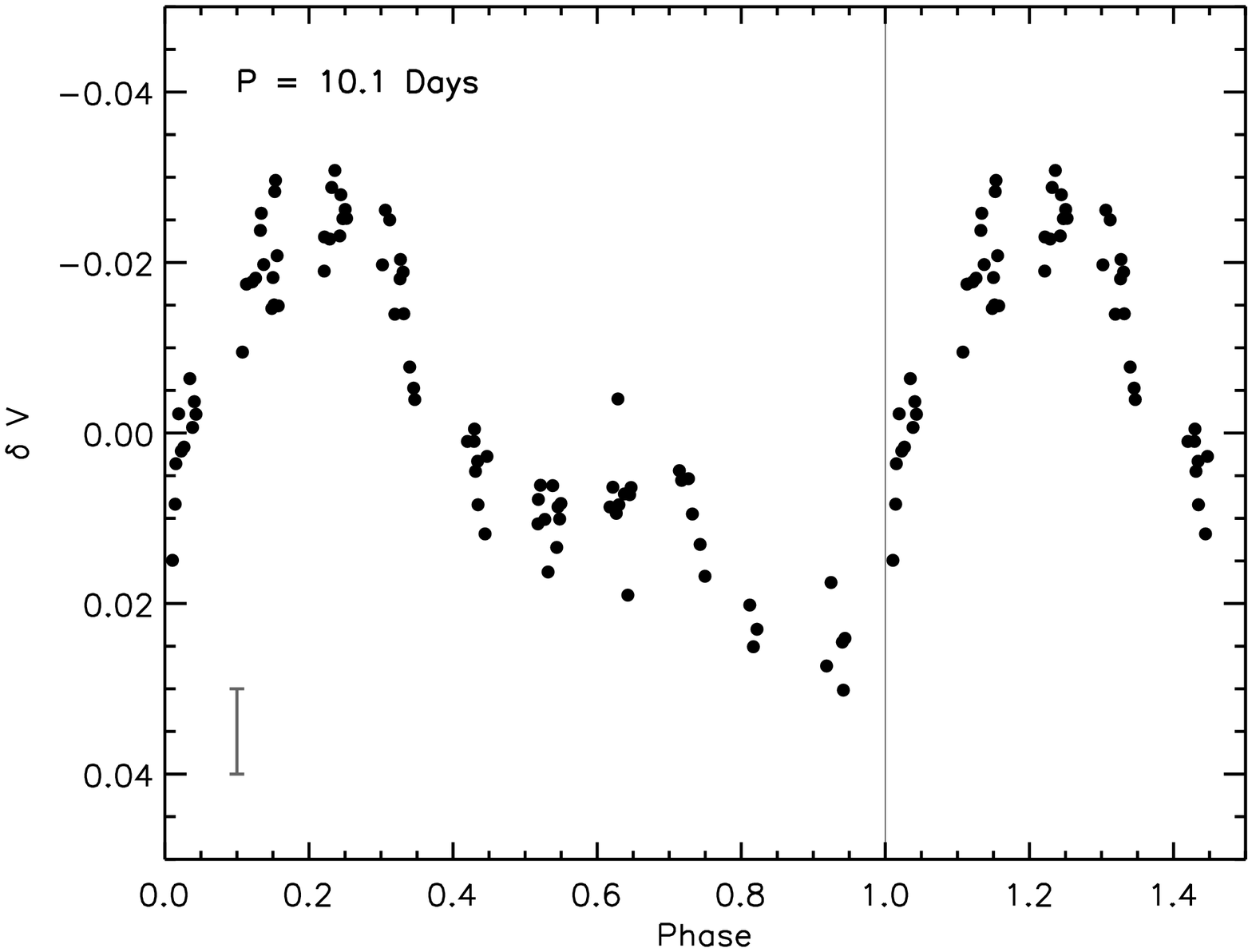}
\caption{{\bf Left:} The light curve for binary 6200 based
on 138 photometric measurements from October 2002 to March 2003.
A sine function with the a 10.13 days period overplotted.
{\bf Right:} The differential V-band photometry for binary 6200
phased to a period of $10.13 \pm 0.39$ days, corresponding to
the maximum periodogram power. The vertical solid line indicates
a phase value of 1.0, and the errorbar in the lower left-hand
corner represent $\pm$ the typical photometric error at
the V magnitude of binary 6200.
\label{6200lc}}
\end{figure}

Figure~\ref{6200lc} shows the light curve and phased
light curve of binary 6200 based on 86 photometric measurements
from December 2002. A total of 138 measurements were
made from October 2002 to March 2003. The maximum
periodogram power corresponds to a period of $10.13
\pm 0.39$ days. The same period was found using all
138 brightness measurements, but the periodic signal
is noisier. We note the structure in the light curve
between phase 0.6 and 0.8. This secondary signal is
presumably due to a second group of photospheric spots.

\begin{figure}[h!]
\epsscale{1.0}
\plotone{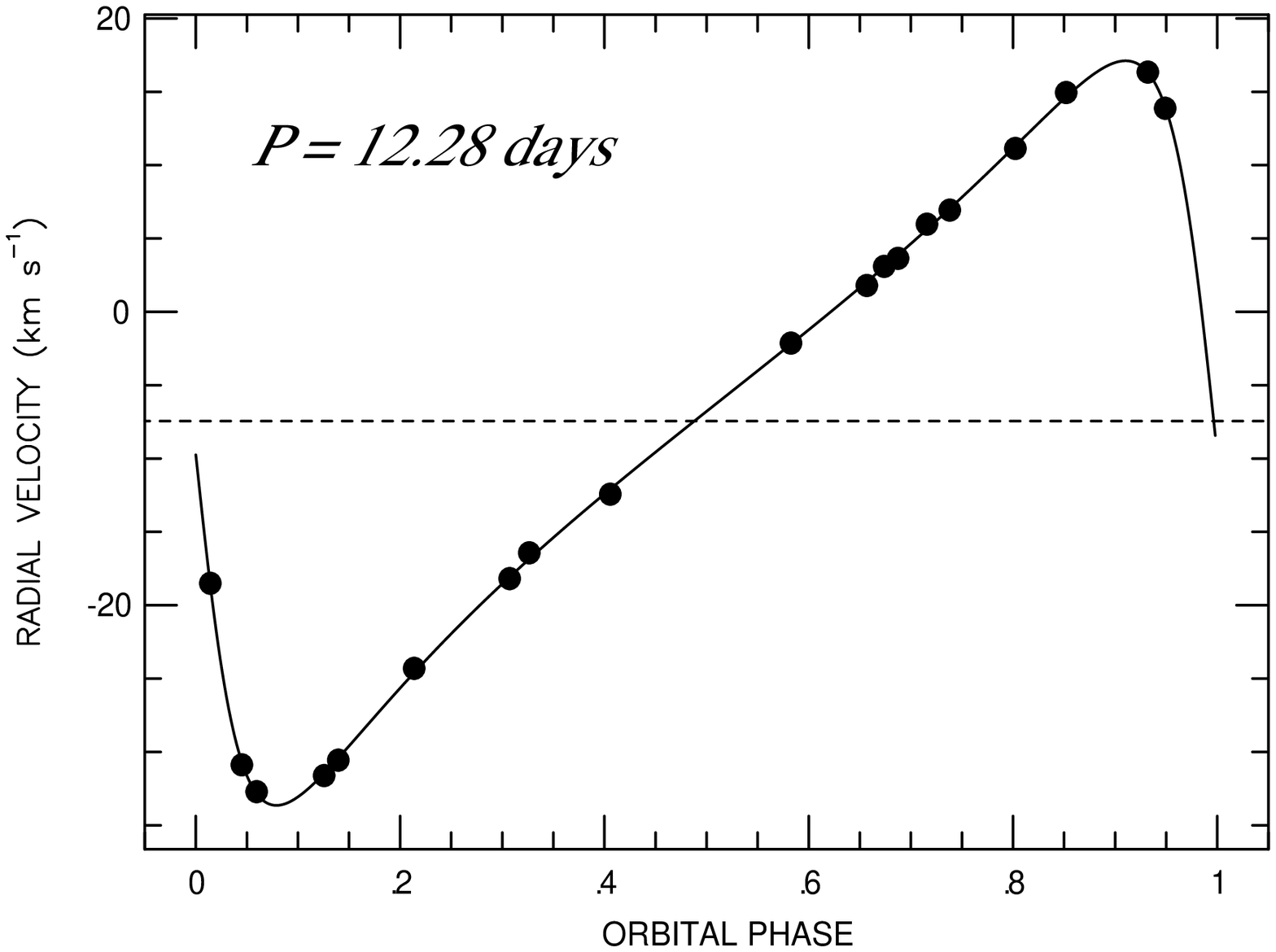}
\caption{The radial velocity measurements of
binary 3081 phased to an orbital period of 12.28 days.
The best fit orbital solution is over-plotted. The eccentricity
of the orbit is $0.550 \pm 0.003$.
\label{3081orb}}
\end{figure}

{\it Binary 3081:}
The orbital solution for binary 3081 is based on 19 radial-velocity
measurements over $\sim$ 70 cycles, and gives an orbital period
of 12.28 days and an orbital eccentricity of $0.550 \pm 0.003$.
Figure~\ref{3081orb} shows the orbital solution over-plotted
on the phased radial velocity data. We estimate the mass of the
primary star to be $\sim 1.1~M_{\odot}$.
The radial-velocity cluster membership probability is 93\%.

\begin{figure}[h!]
\epsscale{1.0}
\plottwo{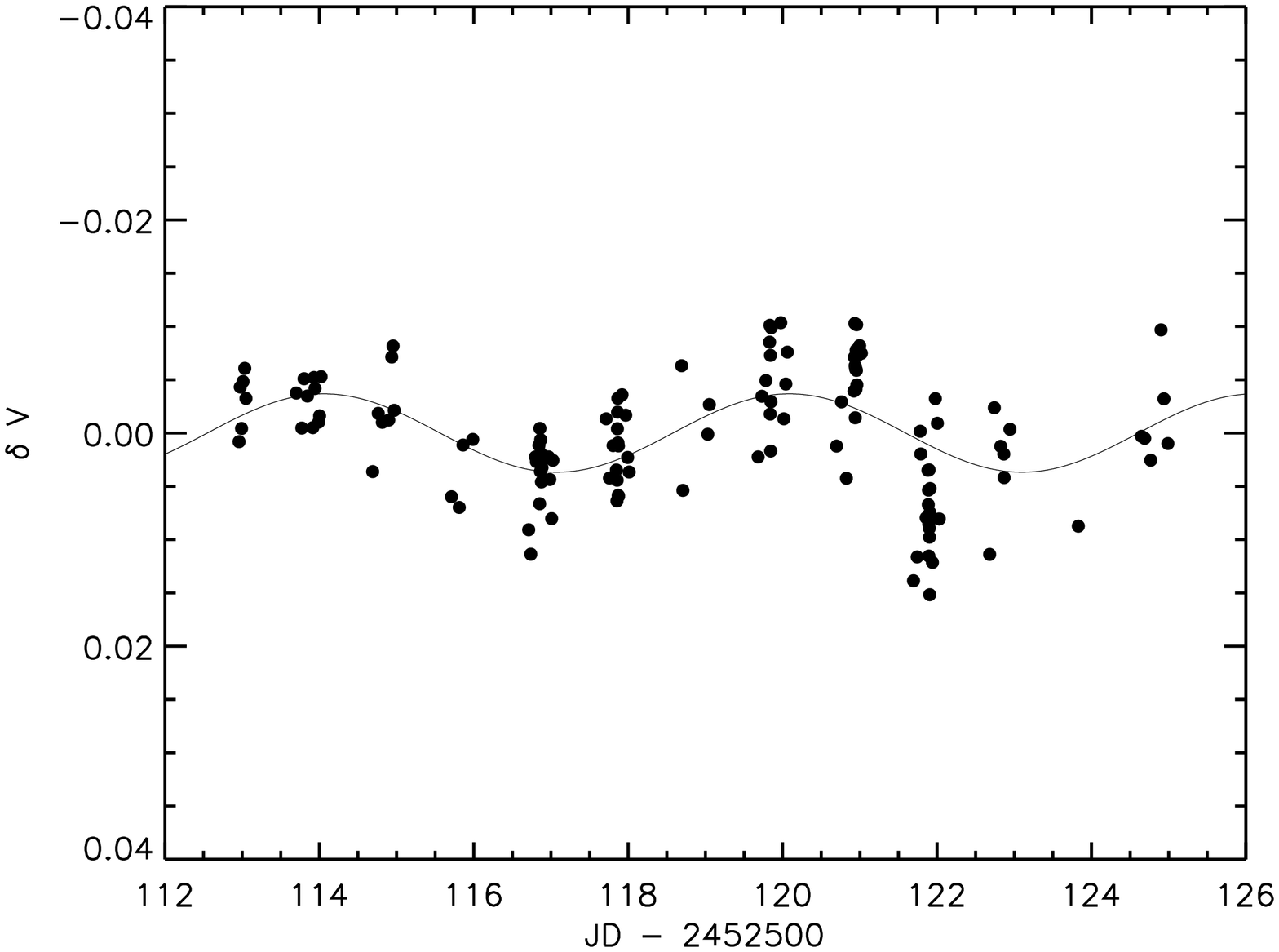}{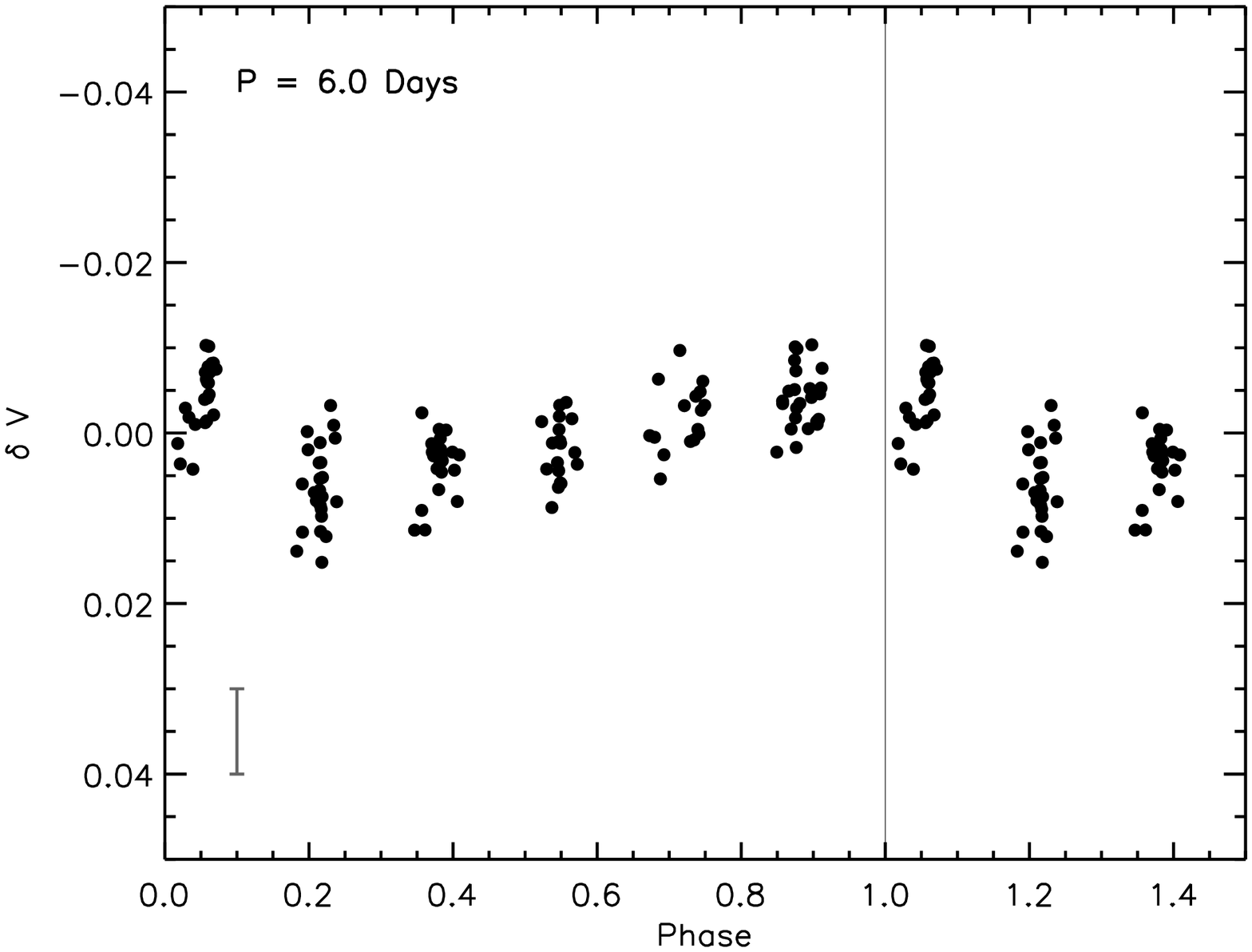}
\caption{{\bf Left:} The light curve for binary 3081 based
on 133 photometric measurements from December 2002. A sine
function with the a 6.03 days period overplotted. {\bf Right:}
The differential V-band photometry for binary 3081 phased to a
period of $6.03 \pm 0.12$ days, corresponding to the maximum periodogram
power. The vertical solid line indicates a phase value of 1.0,
and the errorbar in the lower left-hand corner represent $\pm$
the typical photometric error at the V magnitude of binary 3081.
\label{3081lc}}
\end{figure}

Figure~\ref{3081lc} shows the light curve and phased light curve based
on 133 photometric measurements from December 2002. The rotation
period corresponding to the maximum periodogram power is $6.03
\pm 0.12$ days. The grouping of the data in the phased light
curve is caused by the integer value of the rotation period
and the data sampling frequency. The V magnitude of binary
3081 is $14\fm37$ and its amplitude of variability is $0\fm02$,
approximately four times the expected photometric error (see
Figure~\ref{m0_sigst}).

\section{THE POTENTIAL PHOTOMETRIC EFFECTS OF BINARITY} \label{effects}

Tidal synchronization is driven by dissipation within the star.
Models of synchronization are thus sensitive to the stellar
interior structure. To properly constrain such models, observations
of tidal synchronization must be obtained for stars with known
masses (structure). For single-lined spectroscopic binaries
only the mass of the primary star can be determined.
However, the brightness variations in a binary may be caused by
effects other than spots on the surface of the primary star. We
are not concerned with stellar eclipses as they produce a characteristic
and easily detectable photometric effect that with little difficulty
can be distinguished from spot modulation. Still, other phenomena
may cause photometric variability similar to that of spots on
the primary star. We identify here two potential sources of
photometric variability and estimate the influence of each of
these effects on our ability to determine the rotation period
of the primary stars from spot modulation. 

\subsection{THE EFFECT OF SPOTS ON THE SECONDARY STAR}

Let us first consider the effect of a spot on the surface
of the secondary star. Because all of the binaries presented
here are single-lined spectroscopic binaries, we will assume
that the V band ($\sim 5000$\AA\ ) flux of the secondary star
is at least a factor of five less than that of the primary.
We further assume that the spot on the secondary star produces
an observed peak-to-peak brightness variation of the secondary
of $0\fm15$ in the V-band, equivalent to the largest periodic
signals observed in M35. The spot on the secondary, by itself,
will then result in a $0\fm02$ peak-to-peak variation in the
brightness of the binary.

The observed peak-to-peak brightness variations of the
binaries presented in this paper are in the range from
$\sim 0\fm02 - 0\fm12$. Therefore, to assign the observed
variability of these binaries to spots on the secondary
will require the combination of a heavily spotted secondary
star (flux reduction of up to 75\%) and a quiet (spot-less)
primary star. Such a combination seems unlikely. The detected
photometric variability is thus likely to result from spots
on the primary star.

If, in a binary, both stellar spins have been synchronized
or pseudo-synchronized to the orbital motion, then photometric
variability at the $\sim 0\fm02$ level can be due to spots on
both stars that appear in phase as seen by the observer.
Arguably, stars in the majority of young solar-type binaries
rotate out of phase and non-synchronous with their eccentric
orbital motion. The variability, if any, in the combined light
of such binaries will derive from an out-of-phase superposition
of the periodic signals caused by spots on both stars. Photometric
time-series studies of solar-type main-sequence stars typically
detect periodic variability at the level of $\sim 0\fm02 - 0\fm2$,
and we know from the results presented here that such photometric
variability is not confined to single stars. Therefore, we argue
that the photometric variability detected in this and other
studies comes from one star, either a single star or the primary
star in a binary system.

\subsection{THE EFFECT OF TIDAL DEFORMATION}

Another potential source of brightness variability 
in a binary star is the change in brightness due to
tidally induced changes in the projected surface area
of the stars. When a star becomes oblate due to the
gravitational forces in the binary systems, its projected
surface area, as seen by an observer, will change as it
revolves. The change in area will depend on the binary
mass ratio, the stellar separation, and on the inclination
of the orbit to the line of sight. For a circularized binary
the resulting brightness variation will be sinusoidal with
a period equal to half the orbital period. For a binary with
an eccentric orbit the brightness will increase at periastron
passage and the light curve will deviate from a sinusoidal
shape.

\citet{zahn92} estimates the elevation, $\delta R$,
of a tide on a star with mass $M$ and radius $R$
raised by a companion with mass $m$ at a distance $d$,
as $\delta R / R \simeq q(R/d)^3$, where $q = m/M$ is
the binary mass ratio. If we assume that $M = 1.0~M_{\odot}$
and $m = 0.7~M_{\odot}$ and that the orbital period
is 2.25 days (as in our shortest period binary). Then
$d = 0.04$ AU and $\delta R \simeq 0.08~R_{\odot}$. To
estimate an upper limit on the consequent photometric
variation we assume that the binary is seen ``edge on''
($i = 90\degr$), and that the radius of the stellar disk
increases/decreases by $\delta R$ when the line joining
the two stars is perpendicular/parallel to the line of
sight. The resulting maximum difference in projected
surface area of both stars corresponds to a brightness
difference of $\sim 0\fm02$ for the binary.

Note that this upper limit estimate for the variability
introduced by tidal distortion is for a stellar separation
corresponding to our shortest period binary. The effect
will decrease rapidly with stellar separation and thus
will be negligible for all binaries other than 1455.
Binary 1455 has a 2.25 day circular orbit and so the
photometric variability due to tidal distortion in this
system should have a period of 1.125 days and an amplitude
of $\sim 0\fm02$. The light curve of 1455 actually varies with a
period of 2.3 days and with an amplitude of $0\fm08$.

We conclude from this analysis that the periodic variability
detected in the light curves of the unevolved single-lined M35
and m34 binaries are caused by spots on the solar-type primary
stars, and are therefore reliable measures of their rotation
periods. Thus, in what follows, we will compare our observations
of tidal synchronization to models using solar-type binary components.


\section{THE $\log(\Omega_{\star}/\Omega_{ps}) - \log(P)$ DIAGRAM} \label{loop}

\begin{figure}[h!]
\epsscale{1.0}
\plotone{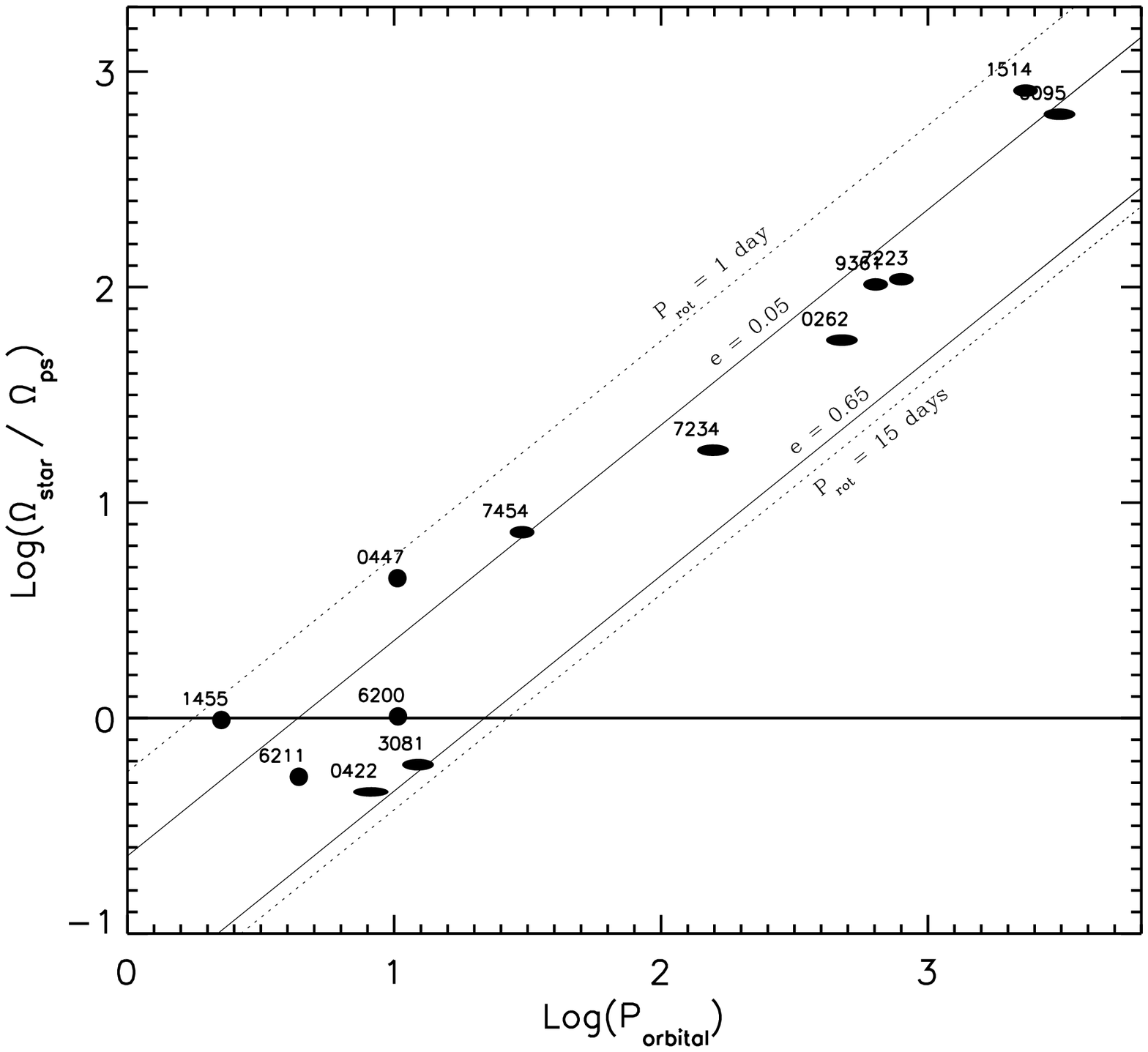}
\caption{The $\log(\Omega_{\star}/\Omega_{ps}) - \log(P)$
diagram for M35 and M34$. \log(\Omega_{\star}/\Omega_{ps})$
is plotted as a function of orbital period for all 13 binaries.
The shape of the plotting symbols are indicative of the
orbital eccentricity of each binary. A solid horizontal
line (the synchronization line) mark $\log(\Omega_{\star}/\Omega_{ps})
= 0$, the location in the diagram for synchronized or
pseudo-synchronized binaries ($\Omega{\star} = \Omega_{ps}$).
$\log(\Omega_{\star}/\Omega_{ps}) > 0$ for super-synchronous
binaries and $\log(\Omega_{\star}/\Omega_{ps}) < 0$ for
sub-synchronous binaries. The interval framed by solid lines
represents $\log(\Omega_{\star}/\Omega_{ps})$ corresponding
to $P_{rot}^{prim} = 4.3$ days and $0.05 < e < 0.65$. The
interval framed by dotted lines represents $\log(\Omega_{\star}/\Omega_{ps})$
corresponding to $e = 0.35$ and $1 < P_{rot}^{prim} < 15$ days.
It is expected that binary primary stars, unaffected by tidal
evolution, will be distributed in the area of the
$\log(\Omega_{\star}/\Omega_{ps}) - \log(P)$ diagram
enclosed by the dotted lines. However, tidal theory predicts
that primaries in binaries with periods similar to or shorter
than the tidal circularization period ($10.2^{+1.0}_{-1.5}$ days)
should be either (pseudo-) synchronized or rotate slightly
super-synchronous and thus fall on or slightly above the
synchronization line. The deviation from synchronization line
of four of the six primaries with $\log(P) \la 1.2$ thus offer
interesting challenges to our understanding of tidal evolution
in solar-type binaries.
\label{loolp}}
\end{figure}

In this section we present our observational results in a way
that facilitates comparison with predictions of tidal theory.
In that spirit, we introduce in Figure~\ref{loolp} the
$\log(\Omega_{\star}/\Omega_{ps}) - \log(P)$ diagram. From our
observational results we can derive for each binary the average
orbital angular velocity ($\omega = 2\pi/P_{orb}$) and the rotational
angular velocities of the primary star ($\Omega_{\star} =
2\pi/P_{rot}^{prim}$). With reference to Section~\ref{theory},
eqs. [3] and [4], we can calculate, for a given binary, the
theoretical pseudo-synchronization angular velocity ($\Omega_{ps}$).
As a diagnostic of the degree of tidal synchronization in a binary
system we use the ratio of $\Omega{\star}$ to $\Omega_{ps}$

\begin{equation}
\frac{\Omega_{\star}}{\Omega_{ps}}~~=~~
\frac{{(1 + 3e^2 + \frac{3}{8}e^4)~(1 - e^2)^{3/2}}}
{1 + \frac{15}{2}e^2 + \frac{45}{8}e^4 + \frac{5}{16}e^6}
\frac{P_{orb}}{P_{rot}^{prim}}
\end{equation}

\noindent The base 10 logarithm of $\Omega{\star}/\Omega_{ps}$
(i.e. $\log(\Omega_{\star}/\Omega_{ps})$) has a useful behavior for
analysis. Synchronous and pseudo-synchronous binaries will lie
on the line represented by $\log(\Omega_{\star}/\Omega_{ps}) = 0$.
We refer to that line as the {\it synchronization line}.
Super-synchronous binaries ($\Omega{\star} > \Omega_{ps}$)
will have $\log(\Omega_{\star}/\Omega_{ps}) > 0$, while
sub-synchronous binaries ($\Omega{\star} < \Omega_{ps}$)
have $\log(\Omega_{\star}/\Omega_{ps}) < 0$. Figure~\ref{loolp}
shows $\log(\Omega_{\star}/\Omega_{ps}$) for all 13 binary
stars as a function of their orbital periods and Table 1
lists the values of $\Omega_{\star}/\Omega_{ps}$ and
$\log(\Omega_{\star}/\Omega_{ps})$. The uncertainties on
both the orbital periods and on $\log(\Omega_{\star}/\Omega_{ps})$
are small and the errorbars fit within the plotting-symbols
for all binaries with periods shortward of 100 days. 
We discuss here the degree of tidal synchronization in
each of the six binaries with orbital periods less than 13 days.

{\it Binary 1455}:
The similarity of the orbital period (2.25 days) and the
rotation period of the primary star ($2.29 \pm 0.02$ days)
suggests that the primary in binary 1455 is synchronized
to the circular orbital motion. Binary 1455 is thus both
tidally circularized and synchronized at the age of 150 Myr,
and lies on the synchronization line in Figure~\ref{loolp}
($\Omega_{\star}/\Omega_{ps} = 0.98$).

{\it Binary 6211}:
This 250 Myr M34 binary has a circular orbit with a period
of 4.39 days but the primary star is rotating sub-synchronously
at $8.03 \pm 0.1$ days, corresponding to 53\% of the orbital
angular velocity ($\Omega_{\star}/\Omega_{ps} = 0.53$, 
$\log(\Omega_{\star}/\Omega_{ps}) = -0.28$).

{\it Binary 0422}:
This 150 Myr M35 binary has not been circularized and has a
highly eccentric orbit ($e = 0.65$) with a period of 8.17 days.
The primary star in 0422 is not pseudo-synchronized and rotates
with a sub pseudo-synchronous period of $3.71 \pm 0.06$ days,
corresponding to 44\% of $\Omega_{ps}$ ($\Omega_{\star}/\Omega_{ps}
= 0.44$, $\log(\Omega_{\star}/\Omega_{ps}) = -0.35$). Due to the
relative brightness of the primary and tertiary star in this system
we cannot exclude the possibility that the detected photometric
variability is due to spots on the tertiary (see discussion in
Section 5.1). In that case, spot-activity on the primary star must
be negligible as the power-spectrum resulting from the photometric
data shows only one peak aside from the typical low-level signal
due to the sampling of the data (the window-function). We assume in
the upcoming discussion that the derived rotation period represents
that of the brightest star in the system, the primary.

{\it Binary 0447}:
The 10.28 day orbit of this M35 binary is circular. The rotation
period of the primary star is super-synchronous at $2.30 \pm 0.02$
days ($\Omega_{\star}/\Omega_{ps} = 4.46$,
$\log(\Omega_{\star}/\Omega_{ps}) = 0.65$).

{\it Binary 6200}:
This M35 binary has a circular 10.33 day orbit and synchronized
primary star. The primary star is rotating once every $10.13 \pm 0.39$
days corresponding to $\Omega_{\star}/\Omega_{ps} = 1.02$ or
$\log(\Omega_{\star}/\Omega_{ps}) = 0.01$.

{\it Binary 3081}:
The primary star of this M35 binary follows a highly eccentric
($e = 0.55$) orbit with a period of 12.3 days, while its rotation
period of $6.03 \pm 0.12$ days correspond to sub-synchronous
rotation ($\Omega_{\star}/\Omega_{ps} = 0.61$,
$\log(\Omega_{\star}/\Omega_{ps}) = -0.21$).

The rotation periods of the primary stars in the M35 binaries
$422$ and $3081$ are approximately half their respective orbital
periods. This result is interesting in light of the known effect
of ``period doubling'' where two spots/spot-groups $\sim 180\deg$
apart on the stellar surface causes the observed period to be half
the true period. While period doubling does occur
\citep[e.g.][]{smm+99,hbm+02}, examinations of the power spectra
and phased light curves do not support doubling of the rotation
periods. We note also that binaries $422$ and $3081$ are highly
eccentric and doubling the rotation periods of the primary stars
will not bring these systems into pseudo-synchronization.

\begin{figure}[h!]
\epsscale{1.0}
\plotone{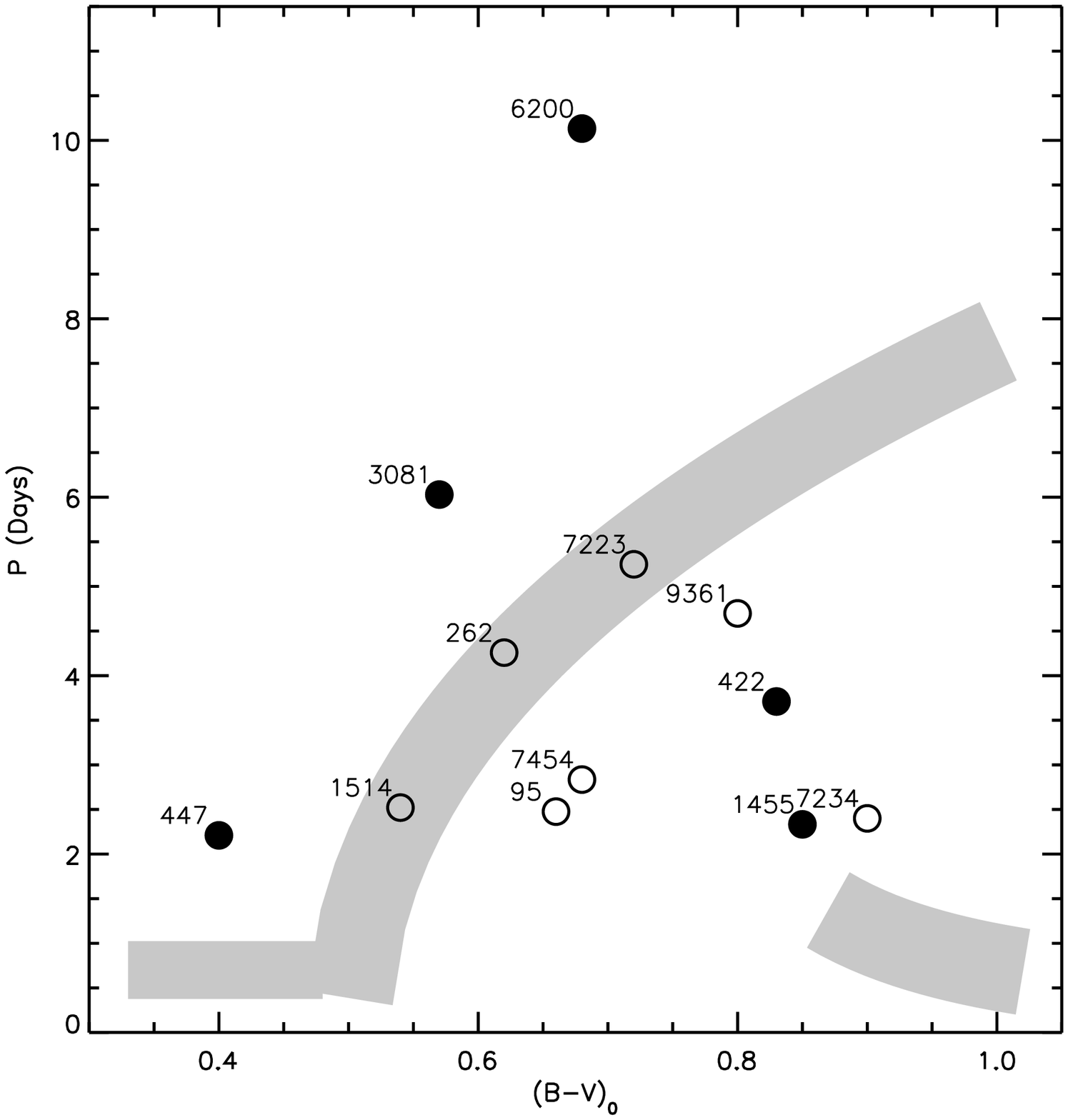}
\caption{The M35 color-period diagram - stellar rotation period plotted
against the stellar color $(B-V)_0$. The broad grey curves represent
well-defined observed sequences of M35 stars. The actual color-period
diagram for M35 will be published in Meibom et al. (2006b). The
locations of the 12 M35 binaries presented in this paper are overplotted
as solid circles (five closest binaries) and open circles (seven wider
binaries).
\label{cpM35}}
\end{figure}


\section{COMPARISON TO THEORETICAL EXPECTATIONS} \label{discussion}

Explicit predictions for tidal evolution based on the equilibrium
or dynamical theories (Section~\ref{theory}) are published for only
a few initial orbital and stellar parameters \citep[see][]{zb89,ws02}.
Ideally, model predictions would exist for a fine grid of such parameters,
allowing for direct comparison with the different binaries observed
in M35 and M34. In the absence of such a detailed theoretical framework
we must compare our observations of tidal evolution to predictions derived
from more general findings of tidal theory. One such finding is the
difference in the timescales for tidal synchronization and tidal
circularization in a given binary system. Primarily due to the
difference between stellar and orbital angular momenta, the timescale
for tidal synchronization is approximately three orders of magnitude
smaller than the timescale for tidal circularization for constant
stellar structure. In lieu of specific theoretical predictions for
the tidal evolution of our binaries,
we use these relative timescales to formulate two simple expectations
for tidal evolution in a coeval sample of {\it main-sequence} binaries:
1) The rotation of a star in a circularized binary should be synchronized
to the orbital angular velocity; and 2) The rotation of a star in an
eccentric binary should be pseudo-synchronized ($\Omega_{\star} =
\Omega_{ps}$) if the orbital period is similar to or shorter than
the tidal circularization period.

While these simple expectations are likely valid for older main-sequence 
stars, the models of \citep{zb89} caution that primary stars in 
binaries as young as those studied here may be supersynchronously 
rotating to a small degree that depends upon the initial stellar and 
binary parameters.

In the $\log(\Omega_{\star}/\Omega_{ps}) - \log(P)$ diagram for M35
and M34, these expectations correspond to all of the shortest period
binaries ($P_{orb} \la 13$ days) being located either on the synchronization
line or slightly above ($\Omega_{\star}/\Omega_{ps} \la 0.2$). Inspection
of Figure 11 immediately shows that this is not the case.

The two shortest period binaries, 1455 in M35 and 6211 in M34, both
have circular orbits, as expected. Furthermore, the primary star of
the shortest period binary, 1455, is indeed synchronized, in agreement
with the first expectation. In marked contrast, the primary star of
6211 is rotating sub-synchronously in a circular orbit, and thus runs
counter to the most basic expectation of main-sequence tidal evolution.
We note that the circular orbit of 6211 is not a surprise; in M34 the 4.4 day
period is less than half that of the tidal circularization period of
the younger M35, and three additional circular binaries have been found
in M34 with periods between 4 and 5.5 days. Thus it is the newly
discovered sub-synchronism that requires explanation.

The M35 binaries 0447 and 6200 provide another important comparison.
Both have circular orbits at essentially the same 10.3 day period.
However, the primary star of 6200 is rotating synchronously, as expected,
while the primary of 0447 is rotating super-synchronously by more than
a factor 4. This high degree of super-synchronism is unexpected based
both on the circular orbit of 0447 and on comparison with 6200, its
near-twin in period, eccentricity, and age. Perhaps of importance is
the difference in the primary masses; $1.4~M_{\odot}$ for 0447 and
$1.1~M_{\odot}$ for 6200. This difference in mass, and thus interior
structure, of the primary stars can potentially translate into a difference
in the mechanism and efficiency for tidal dissipation, in the internal
angular momentum transport, and in the rate of external angular momentum
loss (e.g., wind loss). Binaries 0447 and 6200 thus provide a remarkable
test of the role of stellar mass and structure in tidal evolution.

The third diagnostic pair of binaries are the M35 eccentric binaries
0422 and 3081. 0422 has a period of 8.17 days, shorter than the tidal
circularization period of M35, yet retains an eccentricity of 0.65.
3081 has a period of 12.3 days, slightly longer than the M35 tidal
circularization period, and has an eccentricity of 0.55. Both binaries
represent the best cases in our sample to study evolution to
pseudo-synchronization. In fact, neither have achieved
pseudo-synchronization; both are rotating sub-synchronously by factors
of 2.2 and 1.7. As such they both are counter to the second expectation
of pseudo-synchronism for periods near the tidal circularization period.

The primaries of three of these six binaries are observed to rotate
sub-synchronously. Stellar differential rotation and spots at high
latitudes could lead to observed sub-synchronous rotation not
representative of the rotation at the stellar equator. Using the Sun
as a reference, the difference in rotation period between the equator
(25 days) and a latitude of $60\degr$ (32 days) of a solar-type star
is 7 days. The angular rotation velocity determined from brightness
variations due to a spot at $60\degr$ latitude will thus be about 25\%
less than if the spot were located at the equator. Assuming that this
is the case for the primary stars in binaries 6211, 0422, and 3081,
and thus increasing $\Omega_{\star}$ of the primary stars in these
binaries by 25\%, leads to $\Omega_{\star}/\Omega_{ps}$ ratio of 0.66,
0.75, and 0.57, respectively, for binaries 6211, 3081, and 0422. Such
corrections are not sufficient to bring these binaries into synchronous
or pseudo-synchronous rotation. We conclude from these estimates that
the sub-synchronous rotation observed in the 3 binaries can not be
explained due to differential rotation and spots at high latitude
unless differential rotation with latitude is more severe in younger
stars as compared to the sun.

To summarize, among six young (150-250 Myr), short-period ($< 13$ days),
solar-type binaries, only two have reached the equilibrium state
of both a circularized orbit and synchronized rotation. Among the
exceptions we find binaries with circular orbits that are not synchronized
(one being supersynchronous and one subsynchronous), and binaries
with eccentric orbits that are not pseudo-synchronized (both being
subsynchronous). As a set, these six binaries present a challenging
case study for tidal evolution theory. Here we begin that conversation
with brief considerations of stellar evolution, stellar dynamics,
tidal evolution theory, and initial conditions. None is capable of
explaining all of the binaries.

Given the young ages of M34 and M35, these solar-mass stars have not
been on the main sequence for long; hence the assumption of constant
interior structure likely should be abandoned and the impact of PMS
evolution considered. In fact, super-synchronous rotation
($\Omega_{\star}/\Omega_{ps} \simeq 0.2$) at the age of M35 is predicted
by \citet{zb89} for a circular 7.8 day period binary comprising two
$1~M_{\odot}$ stars. In their model the super-synchronous rotation
on the early main-sequence is a result of a weakened tidal torque
as the convective envelope of the primary retreats during late PMS
evolution. Consequently the primary star is not tidally locked as
its radius decreases, and the star spins up as it evolves to the main
sequence. Perhaps this is the explanation for the super-synchronous
rotation of binary 0447. In this scenario its difference from binary 6200
would be attributed to a smaller convective envelope of the higher
mass primary star in 0447, and thus relatively less tidal coupling
than in 6200 since reaching the ZAMS. We note that the ratio of
$\Omega_{\star}$ to $\omega$ for 0447 is almost three times the
ratio predicted by \citeauthor{zb89} at the age of M35, but their
model calculation is for two solar-mass stars in any case. Indeed,
arguably the synchronization of the $1.1~M_{\odot}$ primary in the
circularized 6200 with a 10.3 day period is a more significant
contradiction with the \citeauthor{zb89} model.

Our findings for the binaries 0422 and 3081 may also reflect on the
dynamical evolution of the cluster. M35 is a rich cluster with a
half-mass relaxation time of order 100 Myr years (Mathieu 1983).
It is likely that some binaries in the cluster core have gone through
resonant gravitational encounters and stellar exchanges. Typically
such encounters lead to hard, highly eccentric orbits for the binary
products, and so perhaps the eccentric binaries 0422 and 3081 are
such products. If so, the most massive stars - i.e., the primary stars
- are the most likely to have been exchanged into the binary. As such,
they would have experienced tidal torques for a time smaller than the
age of the cluster, and indeed may retain their rotation periods from
the time of the encounters. In this scenario, that both are sub-synchronous
is essentially the result of chance selection of those rotation
periods from the single star population.

Finally, we note that the null hypothesis that the rotation periods
of the primary stars are not at all influenced by tidal effects cannot
be definitively ruled out. We show in Figure~\ref{loolp} two intervals,
one bounded by diagonal dotted lines and the other bounded by diagonal
solid lines. The interval framed by solid lines represents
$\log(\Omega_{\star}/\Omega_{ps}$) corresponding to $P_{rot}^{prim} =
4.3$ days (the median observed rotation period for stars in M35;
\citet{mms06b}) for $0.05 < e < 0.65$ (the observed range of non-zero
orbital eccentricities in M35 \citep{mm05}). The interval framed by
dotted lines represents $\log(\Omega_{\star}/\Omega_{ps}$) corresponding
to $e = 0.35$ (the mean of the observed eccentricity distribution in
M35) and $1 < P_{rot}^{prim} < 15$ days (the range in rotation periods
in M35). All of the M34 and M35 primary stars with measured rotation
periods fall within these intervals, and thus are not distinguishable
from the rotation periods of single stars. For the binaries with periods
greater than 30 days, we presume that the rotation periods of the primaries
in fact are set by the same mechanisms as in the single stars.

Might the same be said for the six binaries with periods of less than
30 days? The low rotation periods of the primary stars in both of the
eccentric binaries 0422 and 3081 might derive simply from the same
causes of such periods in single stars. In fact, the rate of tidal
evolution in high eccentricity binaries may be slower than for binaries
with low eccentricities. In a highly eccentric orbit tidal interactions
are essentially confined to a short time interval around periastron passage.
If the duration of the tidal perturbation at periastron passage is shorter
than the convective turnover time, the efficiency of tidal coupling may
be significantly reduced (\citet{go97}, \citet{gm91}, and \citet{zahn89}).
\citet{dmm92} and \citet{mm05} conjecture that eccentric binaries with
periods at or below the tidal circularization period, such as 0422 and
3081, might be the result of high primordial eccentricities ($e \ga 0.7$).
Perhaps the consequently reduced rate of tidal evolution has allowed
the rotational evolution of the primary stars to be little different
from that of single stars.

We find it far less plausible that the rotational evolutions of the
primaries of the four binaries with circular orbits have not been
altered by tidal effects. Two of these binaries ($1455$, $6200$)
would have to fall on the synchronization line by chance, while the
circular orbits of the other two ($6211$, $447$) are strong evidence
for substantial action by tidal forces since the formation of the system.
In fact, evidence that tidal synchronization has affected the rotational
evolution of the closest binaries can be found in the ``color-period''
diagram for M35. Figure~\ref{cpM35} shows the M35 color-period diagram;
the stellar rotation periods plotted against their colors
\citep[see also][]{barnes03a}. For clarity in this context, broad
grey curves are plotted to represent well-defined observed sequences
of M35 members (the actual M35 color-period diagram can be found in
\citet{thesis} and will be published in Meibom et al. (2006b))
The locations of the 12 M35 binaries discussed in this paper are
overplotted. The primary stars in binaries $447$, $3081$ and $6200$
rotate abnormally slow with periods close to twice that of similar
M35 stars while wider binaries (open circles) fall on or close to
the sequences representing the typical M35 star. In the cases of
binaries $447$ and $6200$ (synchronized), this abnormally slow rotation
provide further evidence that tidal synchronization has affected their
rotational evolution. 

The observed rotational and orbital states of the six close M35 and M34
binaries pose interesting and different challenges to current tidal theory.
To properly test tidal theory, models must be run with stellar and binary
parameters tailored to fit the observed systems. Furthermore, ingredients
such as internal stellar angular momentum transport and external wind loss
should be included in such models, as both helioseismic observations of the
Sun \citep[e.g.][]{gdk+91,ekj02} and observations of rotation of late-type
stars in general (e.g. \citet{barnes03a,ssm+93}) suggest that angular
momentum is transported between the convective envelopes and the radiative
cores in such stars.


\section{SUMMARY AND CONCLUSIONS} \label{conclusions}

Tidal forces in close detached binaries drive an exchange
of angular momentum between the stars and their orbital
motions. With time a close binary system will approach an equilibrium
state in which the stellar spin axes are aligned perpendicular
to the orbital plane, the stellar spins are synchronized to
the orbital motion, and the stellar orbits are circular.

Tidal theory makes predictions about rates of tidal evolution
in close solar-type binaries. However, current models
cannot account for the extent of tidal circularization observed
in the oldest populations of solar-type binaries, indicating
that the theoretical rates of tidal evolution are too small.
The same models predict that the process of tidal synchronization
proceeds faster than tidal circularization by about three orders
of magnitude. Observations of tidal synchronization therefore
provide an important additional constraint on these models and
the dissipation mechanisms they employ. Importantly, observing
the rate of tidal synchronization also promises to shed light
on physical processes of stars such as internal and external
angular momentum transport. 

We present rotation periods for the solar-type primary stars
in 13 single-lined binaries with known orbital periods and
eccentricities. All 13 binaries are radial-velocity and photometric
members of the young open clusters M35 (150 Myr) and M34 (250 Myr).
The stellar rotation periods are derived from high-precision (0.5\%)
relative time-series photometry obtained from two weeks of classically
scheduled observations combined with $\sim$ 6 months of queue-scheduled
monitoring.

We compare the rotational angular velocity of each primary star
($\Omega_{\star}$) to the angular velocity required for the star
to be (pseudo-) synchronized ($\Omega_{ps}$). We use the value of
$\Omega_{\star}/\Omega_{ps}$ as a measure of the degree of tidal
synchronization and present that measure as a function of the binary
orbital period (P) in a $\log(\Omega_{\star}/\Omega_{ps}) - \log(P)$
- diagram.

Our previous studies of tidal circularization in these clusters
have shown that the orbits of binaries with periods of $\sim$
10 days and less have been altered by tidal interactions.
Considering theoretical predictions that the rate of tidal
synchronization exceeds that of tidal circularization by of
order a factor $10^{3}$ for constant stellar interior structure,
in the context of constant stellar structure we would expect
that 1) circularized binaries would also be synchronized; and
2) the shortest period eccentric binaries would be pseudo-synchronized.
In the case of the solar-type stars in these two young clusters,
the primaries have only recently reached the ZAMS, and one set
of tidal evolution models predict such stars to still be rotating
supersynchronously as a consequence of their PMS
contraction. Thus general theoretical considerations and one set
of specific models lead to the expectation that all binaries with
periods shortward of $\sim$ 10 days would fall on or slightly
above the synchronization line ($\log(\Omega_{\star}/\Omega_{ps})
= 0$) in the $\log(\Omega_{\star}/\Omega_{ps}) -\log(P)$ - diagram.

However, four of the six binaries in M35 and M34 with orbital
periods less than $\sim$ 13 days offer interesting challenges
to our understanding of tidal evolution in solar-type binaries:

\begin{itemize}

\item The 10.28 day orbit of M35 binary 0447 has been circularized,
but the $1.4~M_{\odot}$ primary star rotates highly super-synchronously.
As an important comparison, the M35 binary 6200 ($M_{prim} \simeq
1.1~M_{\odot}$ has been both circularized and synchronized at an
orbital period of 10.33 days.

\item Both primary stars in the highly eccentric M35 binaries
0422 ($M_{prim} \simeq 0.9~M_{\odot}$, $P_{orb} = 8.17$ days)
and 3081 ($M_{prim} \simeq 1.1~M_{\odot}$, $P_{orb} = 12.28$ days)
are rotating slower than their pseudo-synchronization speeds.

\item Orbiting in a 4.39 day circular orbit, the $0.7~M_{\odot}$
primary star in the M34 binary 6211 is also rotating sub-synchronously.

\end{itemize}

\noindent Only binaries 1455 and 6200 meet the expectations of
tidal theory of synchronized primary stars in circular orbits.
Nevertheless, the 10.33 day circularized orbit of binary 6200 
contradicts models predictions of tidal circularization at 150
Myr \citep{mm05}.

At the present time, theoretical models make detailed predictions
for only a few configurations of binary orbital and stellar
parameters. Specific models must be run with carefully chosen
initial orbital and stellar parameters to attempt to reproduce
the observed tidal evolution at 150 Myr and 250 Myr. Furthermore,
future models of tidal evolution will face the challenges of
incorporating the effects of internal and external angular momentum
transport and perhaps the combined effects of the dynamical and
equilibrium tides in late-type stars.

We propose explanations for the observed binary and stellar
parameters within the framework of current theory of stellar
and tidal evolution, stellar dynamics, and observed stellar
and binary initial conditions. Specifically we suggest that
the super-synchronous rotation of binary 0447 might be explained
by PMS and main-sequence tidal evolution \citep{zb89} and the
relatively high mass and shallow convection zone of the primary
star. We also note that that the sub-synchronous rotation of the
primary stars in binary 0422 and 3081 might be due to either
dynamical stellar interactions and/or reduced tidal dissipation
in highly eccentric systems. We offer no explanation of the
sub-synchronous and circular binary 6211, but find it unlikely
that the parameters of binary 6211 together with the three other
circularized binaries are the result of chance initial conditions.
Sub-synchronous rotation has been predicted in close solar-type
main-sequence binaries when loss of stellar angular momentum due
to magnetic-wind braking is considered \citep{zahn94}. However,
only circular binaries were considered and the predicted levels
of sub-synchronism were much smaller than observed here.
 
Populating the $\log(\Omega_{\star}/\Omega_{ps})-\log(P)$
diagram with coeval homogeneous populations of binary stars
sets the beginning of a new era in observational studies
of tidal synchronization. With time, the $\log(\Omega_{\star}/\Omega_{ps})
-\log(P)$ diagram for binary populations spanning in age
from the PMS to the late main-sequence phase will become
an important observational tool tracing the evolution of
tidal synchronization in a way similar to the $e-\log(P)$
diagram in studies of tidal circularization. While at this
early time, the limited number of binaries that can be placed
in the diagram does not allow us to determine a ``tidal
synchronization period'' marking the transition between
synchronous and asynchronous systems, the degree of tidal
synchronization in individual binaries provide interesting
challenges to tidal theory. The success of theoretical
models can be measured by their ability to predict the
observed orbital and rotational evolution of these binary
stars.

\acknowledgments

We are grateful to the University of Wisconsin - Madison
for the time granted on the WIYN 0.9m and 3.5m telescopes.
We would like to express our appreciation for exceptional and
friendly support of site managers and support staff at both telescopes.
We are thankful to all observers in the WIYN 0.9m consortium who provided
us with high-quality data through the synoptic observing program.
We thank Dr. Sydney Barnes for making the photometric measurements
of binary 6211 available to us, Dr. Imants Platais for providing
2MASS ID's, and our colleagues at the 3rd Granada workshop on Stellar
Structure Tidal Evolution and Oscillations in Binary Stars for fruitful
discussions. We thank the referee for a careful review of the paper
resulting in several insightful recommendations that strengthened
the paper. This work has been supported by NSF grant AST 97-31302
and by a Ph.D fellowship from the Danish Research Agency (Forskningstyrelsen)
to S.M.

\newpage




\end{document}